\documentclass[aps,pre,nobibnotes,nofootinbib,showpacs,reprint]{revtex4-1}

\usepackage{graphicx}
\usepackage{amsmath} 	

\newcommand{\disp}[1]{$\displaystyle{#1}$}

\newcommand{\vect}[1]{\mathbf{#1}}
\newcommand{\svect}[1]{\mbox{\boldmath $#1$\unboldmath}}
\newcommand{\eps}{\epsilon_0}
\newcommand{\mode}{{\veck,\lambda}}
\newcommand{\modep}{{\veck',\lambda'}}
\newcommand{\submode}{_{\veck,\lambda}}
\newcommand{\submodep}{_{\veck',\lambda'}}
\newcommand{\ttheta}{\tilde{\theta}}
\newcommand{\rphase}{\tilde{\theta}({\mode})}
\newcommand{\rphasep}{\tilde{\theta}({\modep})}
\newcommand{\subrphase}{\tilde{\theta}_{\submode}}
\newcommand{\subrphasep}{\tilde{\theta}_{\submodep}}

\newcommand{\avg}[1]{\left\langle{#1}\right\rangle}
\newcommand{\ravg}[1]{{\left\langle{#1}\right\rangle}_{\tilde{\theta}}}

\newcommand{\wo}{\omega_0}
\newcommand{\infint}{\int_{-\infty}^{+\infty}}
\newcommand{\Eq}[1]{Eq.~(\ref{#1})} 
\newcommand{\Fig}[1]{Fig.~\ref{#1}}
\newcommand{\Sec}[1]{Sec.~\ref{#1}}
\newcommand{\App}[1]{Appendix~\ref{#1}}
\newcommand{\pol}{\svect{\varepsilon}}
\newcommand{\Bpol}{\svect{\xi}}
\newcommand{\submodeup}{_{\veck_i,1}}
\newcommand{\submodedown}{_{\veck_i,2}}
\newcommand{\veck}{\vect{k}}
\newcommand{\xs}{x_0}
\newcommand{\Es}{E_0}
\newcommand{\ts}{t_0}

\newcommand{\kdr}{\veck\cdot\vect{r}}
\newcommand{\wt}{\omega t}
\newcommand{\kdrp}{\veck'\cdot\vect{r}}
\newcommand{\wtp}{\omega' t}
\newcommand{\dw}{\Delta\omega}
\newcommand{\dkapa}{\Delta\kappa}

\begin{document}


\title{Dynamics Underlying the Gaussian Distribution of the Classical Harmonic Oscillator in Zero-Point Radiation}

\author{Wayne~Cheng-Wei~Huang}
\email{email: u910134@alumni.nthu.edu.tw} 

\author{Herman~Batelaan}
\email{email: hbatelaan2@unlnotes.unl.edu}

\affiliation{Department of Physics and Astronomy, University of Nebraska-Lincoln, Lincoln, Nebraska 68588, USA }

\begin{abstract}
	 In the past decades, Random Electrodynamics (also called Stochastic Electrodynamics) has been used to study the classical harmonic oscillator immersed in the classical electromagnetic zero-point radiation. Random Electrodynamics (RED) predicts an identical probability distribution for the harmonic oscillator compared to the quantum mechanical prediction for the ground state. Moreover, the Heisenberg minimum uncertainty relation is also recovered with RED. To understand the dynamics that gives rise to this probability distribution, we perform an RED simulation and follow the motion of the oscillator. This simulation provides insight in the relation between the striking different double-peak probability distribution of the classical harmonic oscillator and the Gaussian probability distribution of the RED harmonic oscillator. 
	
	A main objective for RED research is to establish to what extent the results of quantum mechanics can be obtained. The present simulation method can be applied to other physical systems, and it may assist in evaluating the validity range of RED. 
\end{abstract}


\maketitle 


\section{Introduction}

\begin{figure*}
\centering
\scalebox{0.55}{\includegraphics{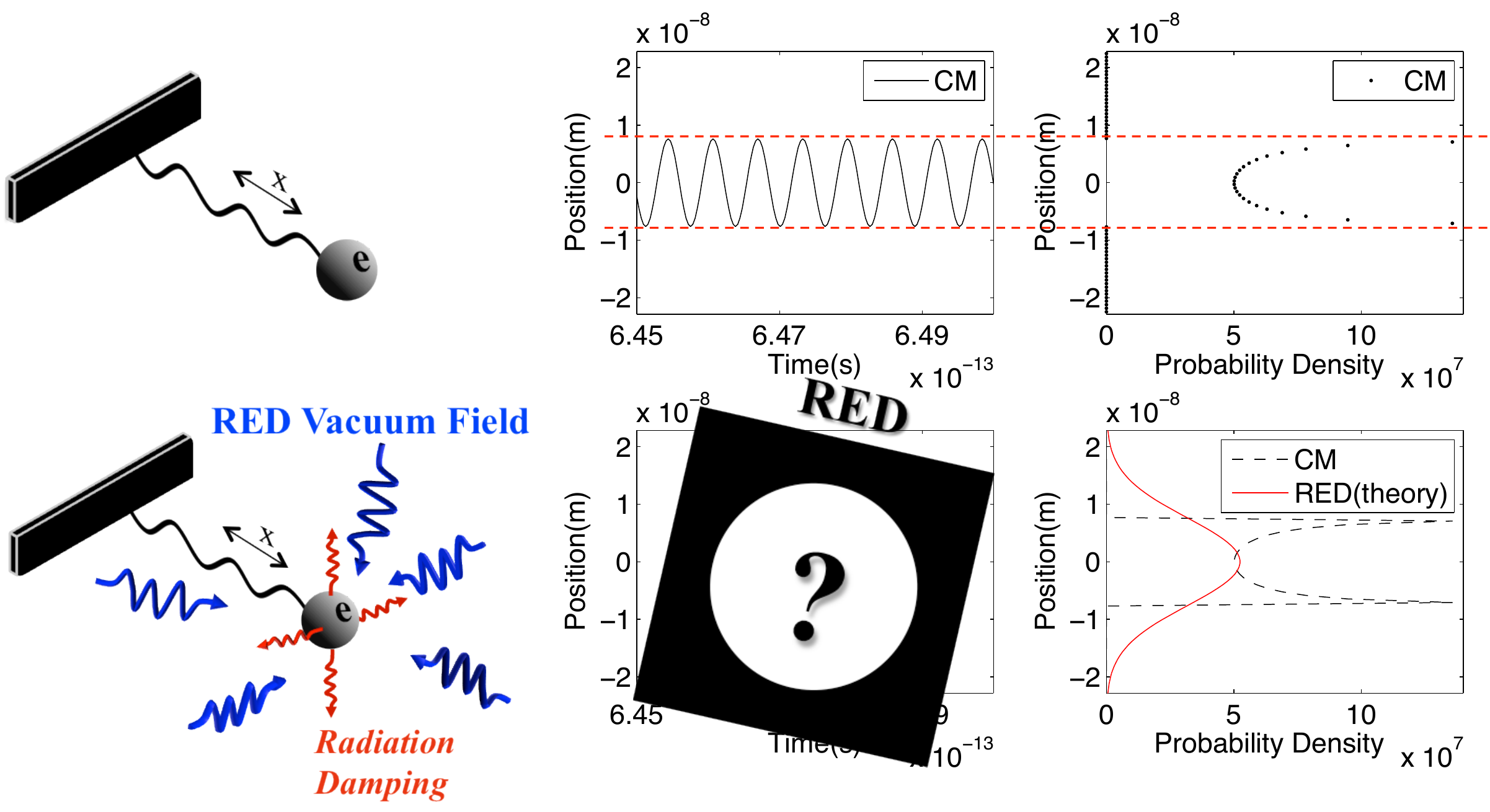}}
\caption{A comparison between the classical and the RED harmonic oscillators. Top: Without any external force, the classic harmonic oscillator that is initially displaced from equilibrium performs a simple harmonic oscillation  with constant oscillation amplitude. The resulting probability distribution has peaks at the two turning points. Bottom: Without any external force except for the RED vacuum field, the RED harmonic oscillator undergoes a motion that results in a Gaussian probability distribution. This motion and its relation to its classical counterpart is investigated in detail in this paper.}
\label{fig:CMREDintro}
\end{figure*}

	According to quantum electrodynamics, the vacuum is not a tranquil place. A background electromagnetic field called the zero-point field, or vacuum field, is always present, independent of any external electromagnetic source \cite{Milonni}. This has inspired an interesting modification to classical mechanics, called random electrodynamics (RED) \cite{Boyer:RED}. As a variation of classical electrodynamics, RED adds a classical zero-point electromagnetic field with random phases to the original theory. In this paper, the name ``RED vacuum field" will be used for this classical electromagnetic field. The RED vacuum field has some properties identical to the QED vacuum field\footnote{
However, the RED vacuum field does not fully correspond to the vacuum field derived in quantum electrodynamics, as discussed in \cite{Milonni, Ibison}.} such as the Lorentz-invariant spectral energy density \cite{Milonni, Boyer:RED}. With the aid of this background field, RED is able to reproduce a number of results that were originally thought to be pure quantum effects \cite{Milonni, Boyer:RED, Boyer:diamagnetism, Boyer:unrah, Boyer:casimir, Puthoff}. In particular, Boyer has shown that the moments $\langle x^n \rangle$ of a harmonic oscillator immersed in the RED vacuum field, called the RED harmonic oscillator, are identical to those of the quantum harmonic oscillator in the ground state \cite{Boyer:RED&QM}. As a consequence, the Heisenberg minimum uncertainty relation is satisfied for an RED harmonic oscillator, and the probability distributions of the RED harmonic oscillator is also the same as that of the ground state quantum harmonic oscillator.

	Although classical mechanics and RED are both theories that gives trajectories of particles, the position probability distributions of the harmonic oscillator in both theories are dramatically different. In classical mechanics, it is most likely to find the particle at the two turning points of the trajectory. Consequently, the probability distribution has two peaks as shown in \Fig{fig:CMREDintro}. On the other hand, the probability distribution for an RED harmonic oscillator is a Gaussian distribution. What happens to the dynamics of the classical harmonic oscillator in the presence of the RED vacuum field such that its probability distributions would become Gaussian? Why do the widths of these distributions satisfy  Heisenberg's minimum uncertainty relation? In this paper, we perform an RED simulation that answers these questions. 	
	
	Most of the results obtained with the RED theory reproduce quantum mechanical results analytically. Whereas Cole has a series of numerical studies on the hydrogen atom \cite{Cole:grd_1, Cole:grd_2, Cole:exct_1, Cole:exct_2}, numerical studies in this field are rare. The major challenge for RED simulation is to incorporate the RED vacuum field which has an unbounded spectrum. A representative selection of vacuum field modes is thus the key to successful RED simulations. In this study, our approach for the selection of modes has been extensively tested for convergence, and we document the details of our simulation so it can be used by others.
	
	The organization of this paper is the following. First, in \Sec{sec:Theory} Boyer's results on the RED harmonic oscillator are briefly reviewed. Based on these results, the probability distribution for the RED harmonic oscillator is derived. Second, in \Sec{sec:RED simulation} a detailed numerical method for simulating the RED vacuum field and the RED harmonic oscillator is presented. Third, in \Sec{sec:Results} the trajectory of the RED harmonic oscillator is solved numerically, and the constructed probability distribution is compared to the analytical probability distribution. Two sampling methods are used in constructing the probability distribution from the simulated trajectories. The first approach is ``sequential sampling'', which is suitable for studying the relation between the dynamics of the RED harmonic oscillator and its probability distribution. The second approach is ``ensemble sampling'', which lends itself well for parallel computing and is convenient for statistical interpretations. Lastly, in \Sec{sec:Discussion} we discuss some potential applications of RED simulation in studying other quantum phenomena, including the electron doubleslit diffraction. Numerical studies have an advantage over the analytical solutions in that they can be adopted to a range of physical systems, and we hope that the current method of simulation may assist in assessing RED's validity range.

\section{Theory}\label{sec:Theory}

 \subsection{Brief Review of Boyer's Work}

	In his 1975 papers \cite{Boyer:RED,Boyer:RED&QM}, Boyer calculated the statistical features of an RED harmonic oscillator, and the Heisenberg minimum uncertainty relation is shown to be satisfied for such an oscillator. The RED vacuum field used in Boyer's work arises from the homogeneous solution of Maxwell's equations, which is chosen to be zero in classical electrodynamics \cite{Boyer:RED}. In unbounded space, the RED vacuum field has an integral form\footnote{
A detailed account of the RED vacuum field in unbounded space is given in \App{app:FreeSpaceField}.},
\begin{gather}\label{Boyer:vacE}
\begin{split}
\vect{E}_{vac}(\vect{r},t) =& \sum_{\lambda=1}^2 \int \! d^3k \; \pol(\mode) \frac{\eta(\mode)}{2} \left( a(\mode)e^{i(\kdr-\wt)} \right. \\
				       & + \left. a^{\ast}(\mode)e^{-i(\kdr-\wt)} \right), \\
\end{split}\\
\eta(\mode)\equiv \sqrt{\frac{\hbar\omega}{8\pi^3\eps}}, \nonumber \\
a(\mode) \equiv e^{i\rphase}, \nonumber
\end{gather}
where $\omega = c|\veck|$, and $\rphase$ is the random phase uniformly distributed in $[0,2\pi]$. The integral is to be taken over all $\veck$-space. The two unit vectors, $\pol(\veck,1)$ and $\pol(\veck,2)$, describe a polarization basis in a plane that is perpendicular to the wave vector $\veck$,
\begin{equation}\label{pol:cond_1}
\pol(\veck,\lambda) \cdot \veck = 0.
\end{equation}
Furthermore, the polarization basis vectors are chosen to be mutually orthogonal,
\begin{equation}\label{pol:cond_2}
\pol(\veck,1) \cdot \pol(\veck,2) = 0.
\end{equation}
To investigate the dynamics of the RED harmonic oscillator, Boyer used the dipole approximation,
\begin{equation}\label{dipoleapprox}
 \kdr \ll 1,
\end{equation}
to remove the spatial dependence of the RED vacuum field, \Eq{Boyer:vacE}, from the equation of motion. Therefore, the equation of motion for an RED harmonic oscillator used in Boyer's analysis is
\begin{gather}\label{EOM:Boyer}
m\ddot{x} = -m\wo^2 x + m\Gamma\dddot{x} + eE_{vac}^{(x)}(t) , \\
\Gamma \equiv \frac{2e^2}{3mc^3} \frac{1}{4\pi\eps}, \nonumber
\end{gather}
where $m$ is the mass, $e$ is the charge, $\wo$ is the natural frequency, and $\Gamma$ is the radiation damping parameter. The $x$-component of the RED vacuum field in \Eq{EOM:Boyer} is 
\begin{equation}
\begin{split}
E_{vac}^{(x)}(t) =& \sum_{\lambda=1}^2 \int \! d^3k \; \varepsilon^{(x)}(\mode)\frac{\eta(\mode)}{2} \left( a(\mode)e^{-i\wt} \right. \\
			  & + \left. a^{\ast}(\mode)e^{i\wt} \right),
\end{split}
\end{equation}
and the steady-state solution is obtained as
\begin{equation}\label{Boyer:sol}
\begin{split}
x(t) =&\; \frac{e}{m}\sum_{\lambda=1}^2 \int \! d^3k \; \varepsilon^{(x)}(\mode)\frac{\eta(\mode)}{2} \left( \frac{a(\mode)}{C(\mode)}e^{-i\wt} \right. \\
	&+ \left. \frac{a^{\ast}(\mode)}{C^{\ast}(\mode)}e^{i\wt} \right)
\end{split}
\end{equation}
where $C(\mode) \equiv \left(-\omega^2 + \wo^2 \right) - i\Gamma\omega^3$. Additionally, using the condition of sharp resonance, 
\begin{equation}\label{sharpreson}
\Gamma\wo \ll 1, 
\end{equation}
Boyer further calculated the standard deviation of position and momentum from \Eq{Boyer:sol} by averaging over the random phase $\tilde{\theta}$ \cite{Boyer:RED&QM}, 
\begin{align}
\sigma_x = \sqrt{\ravg{x^2} - \ravg{x}^2} = \sqrt{\frac{\hbar}{2m\wo}}, \label{length} \\
\sigma_p = \sqrt{\ravg{p^2} - \ravg{p}^2} = \sqrt{\frac{\hbar m\wo}{2}}. \label{momentum}
\end{align}
The above result satisfies the Heisenberg minimum uncertainty relation,
\begin{equation}
\sigma_x\sigma_p = \frac{\hbar}{2}.
\end{equation}
From an energy argument, Boyer showed that this uncertainty relation can be derived from a delicate balance between the gain of energy from RED vacuum field driving and the loss of energy through radiation damping \cite{Boyer:RED}.

 \subsection{Position Probability Distribution}\label{sec:RED probability}
 
	Given the knowledge of the moments $\ravg{x^n}$, the Fourier coefficients $F_{\ttheta}(k)$ of the position probability distribution\footnote{An RED probability distribution can be constructed from either data collected at a sequence of time, or data drawn from an RED ensemble. In the RED ensemble, each member is characterized by the random phase $\ttheta$ of the RED vacuum field. In this paper, the probability distribution constructed from the ensemble sampling method is denoted as $P_{\ttheta}(x)$.} $P_{\ttheta}(x)$ can be determined through the relation
\begin{equation}\label{series}
\begin{split}
F_{\ttheta}(k) &= \infint e^{-ikx}P_{\ttheta}(x) \,dx \\
		     &= \sum_{n=0}^{\infty} \frac{(-ik)^n}{n!} \infint x^nP_{\ttheta}(x) \,dx \\
		     &= \sum_{n=0}^{\infty} \frac{(-ik)^n}{n!} \ravg{x^n}.
\end{split}
\end{equation}
Using \Eq{Boyer:sol} and the relation from Boyer's paper \cite{Boyer:RED&QM}
\begin{equation}
\left\{
\begin{array}{l}
\ravg{e^{\pm i(\rphase+\rphasep)}} = 0, \\
\\
\ravg{e^{\pm i(\rphase-\rphasep)}} = \delta_{\lambda',\lambda}\delta^3(\veck'-\veck), \\
\end{array}
\right.
\end{equation}
the moments $\ravg{x^n}$ can be evaluated,
\begin{align}
\ravg{x^{2m+1}} &= 0, \\
\ravg{x^{2m}} &= \frac{(2m)!}{m!2^m}\left( \frac{\hbar}{2m\wo} \right)^m, \label{moment}
\end{align}
where $m$ is a natural number. Consequently, only even-power terms are contributing in \Eq{series}, and the Fourier coefficients $F_{\ttheta}(k)$ are determined,
\begin{equation}
\begin{split}
F_{\ttheta}(k) &= \sum_{m=0}^{\infty} \frac{(-ik)^{2m}}{(2m)!} \avg{x^{2m}} \\
		     &= \sum_{m=0}^{\infty} \frac{1}{m!} \left( \frac{-\hbar k^2}{4m\wo} \right)^m  \\
		     &= e^{-\frac{\hbar}{4m\wo}k^2}.
\end{split}
\end{equation}
Therefore, although not explicitly given, it is implied by Boyer's work \cite{Boyer:RED&QM} that the position probability distribution of the RED harmonic oscillator is
\begin{equation}\label{prob}
\begin{split}
P_{\ttheta}(x) =& \frac{1}{2\pi} \infint e^{ikx}F_{\ttheta}(k) \,dk \\
        		     =& \sqrt{\frac{m\wo}{\pi\hbar}}e^{-\frac{m\wo}{\hbar}x^2},
\end{split}
\end{equation}
which is identical to the position probability distribution of the quantum harmonic oscillator in the ground state\footnote{
This result is consistent with the phase space probability distribution given in Marshall's work \cite{Marshall}.}. 

\section{NUMERICAL SIMULATION}\label{sec:RED simulation}

\subsection{RED Vacuum Field in Bounded Space}\label{sec:RED vacuum field setup}

	A field confined in a space of volume $V$ with zero value boundary condition has a discrete spectrum, and a summation over the wave vectors $\veck$ is required. On the other hand, the field in unbounded space is not subject to any boundary condition, so every wave vector $\veck$ is allowed. Thus, the field in unbounded space takes an integral form \cite{Boyer:RED}, while the field in bounded space takes a summation form \cite{Milonni, Ibison}. In a simulation, it is convenient to write the RED vacuum field in its summation form,
\begin{gather}\label{vacE}
\begin{split}
\vect{E}_{vac} &= \sum_{\mode} \sqrt{\frac{\hbar\omega}{\eps V}} \frac{1}{2} \left(a_{\submode}e^{i(\kdr-\wt)} + a^{\ast}_{\submode}e^{-i(\kdr-\wt)}\right) \pol_{\submode} \\
&= \sum_{\mode} \sqrt{\frac{\hbar\omega}{\eps V}} \cos(\kdr-\wt+\subrphase) \pol_{\submode}, 
\end{split}
\\
a_{\submode} \equiv e^{i\subrphase}, \nonumber
\end{gather}
where $\omega = c|\veck|$, $\subrphase$ is the random phase uniformly distributed in $[0, 2\pi]$, and $V$ is the volume of the bounded space. A derivation of the summation form of the RED vacuum field in bounded space is given in \App{app:BoundedSpaceField}.

	Since the range of the allowed wave vectors $\veck$ is over all $\veck$-space and the RED vacuum field diverges for infinitely large $|\veck|$, we choose to sample only the wave vectors $\veck$ whose frequencies is within the finite range $[\wo-\Delta/2,\wo+\Delta/2]$. Such sampling is valid as long as the chosen frequency range $\Delta$ completely covers the characteristic resonance width $\Gamma\wo^2$ of the harmonic oscillator, 
\begin{equation}\label{resonshell}
\Gamma\wo^2 \ll \Delta.
\end{equation}
On the other hand, the distribution of the allowed wave vectors $\veck$ depends on the specific shape of the bounded space. In a cubic space of volume $V$, the allowed wave vectors $\veck$ are uniformly distributed at cubic grid points, and the corresponding RED vacuum field is
\begin{equation}\label{vacEcubic}
\vect{E}_{vac} =\sum_{\lambda=1}^{2} \sum_{(k_x,k_y,k_z)} \sqrt{\frac{\hbar\omega}{\eps V}} \cos(\kdr-\wt+\subrphase) \pol_{\submode}.
\end{equation} 
The sampling density is uniform and has a simple relation with the space volume $V$,
\begin{equation}\label{dnsty}
\rho_{\veck} = \frac{V}{(2\pi)^3}. 
\end{equation}
Nevertheless, such uniform cubic sampling is not convenient for describing a frequency spectrum. In order to sample only the wave vectors $\veck$ in the resonance region, spherical coordinates are used. In addition, for the sampling to be uniform, each sampled wave vector $\veck$ must occupy the same size of volume element $\Delta^3k = k^2\sin{\theta}\Delta k \Delta \theta \Delta \phi$. To satisfy both conditions, we use a set of specifically chosen numbers $(\kappa_{ijn}, \vartheta_{ijn}, \varphi_{ijn})$ to sample the wave vectors $\veck$. Namely, for $i = 1\ldots N_{\kappa}$, $j = 1\ldots N_{\vartheta}$, and $n = 1\ldots N_{\varphi}$,
\begin{equation}\label{mslct:spher1}
\left\{ 
\begin{array}{l}
\kappa_{ijn} = (\wo-\Delta/2)^3/3c^3 + (i-1)\dkapa \\
\vartheta_{ijn} = -1 + (j-1)\Delta\vartheta \\
\varphi_{ijn} = R_{ij}^{(0)} + (n-1)\Delta\varphi, 
\end{array}
\right.
\end{equation}
where $R^{(0)}$ is a random number uniformly distributed in $[0,2\pi]$, and the stepsizes are constant,
\begin{align}
\dkapa &= \frac{\left[ (\wo+\Delta/2)^3/3c^3 - (\wo-\Delta/2)^3/3c^3 \right]}{N_{\kappa}-1}, \label{dka} \\
\Delta\vartheta &= \frac{2}{N_\vartheta-1}, \\
\Delta\varphi &= \frac{2\pi}{N_\varphi}.
\end{align}
Such a set of numbers $(\kappa_{ijn}, \vartheta_{ijn}, \varphi_{ijn})$ is then used for assigning the spherical coordinates to each sampled wave vector $\veck$,
\begin{equation}\label{mslct:spher2}
\veck_{ijn}
= 
\begin{pmatrix}
k^{(x)}_{ijn} \\ k^{(y)}_{ijn} \\ k^{(z)}_{ijn} 
\end{pmatrix}
=
\begin{pmatrix}
k_{ijn}\sin{(\theta_{ijn})}\cos{(\phi_{ijn})} \\
k_{ijn}\sin{(\theta_{ijn})}\sin{(\phi_{ijn})} \\
k_{ijn}\cos{(\phi_{ijn})}
\end{pmatrix},
\end{equation}
where
\begin{equation}\label{mslct:spher3}
\left\{ 
\begin{array}{l}
k_{ijn} = (3\kappa_{ijn})^{1/3} \\
\theta_{ijn} = \cos^{-1}(\vartheta_{ijn}) \\
\phi_{ijn} = \varphi_{ijn}.
\end{array}
\right.
\end{equation}
Consequently, each sampled wave vector $\veck_{ijn}$ will be in the resonance region and occupy the same size of volume element,
\begin{equation}
\begin{split}
\Delta^3k &= k^2\sin{\theta}\Delta k \Delta \theta \Delta \phi \\
	        &= \dkapa \Delta \vartheta \Delta \varphi \nonumber.
\end{split}
\end{equation}
Under the uniform spherical sampling method (described by Eqs.~(\ref{mslct:spher1}), (\ref{mslct:spher2}), and (\ref{mslct:spher3})), the expression for the RED vacuum field, \Eq{vacE}, becomes
\begin{equation}\label{vacEspherical}
\vect{E}_{vac} =\sum_{\lambda=1}^{2} \sum_{(\kappa, \vartheta, \varphi)} \sqrt{\frac{\hbar\omega}{\eps V}} \cos(\kdr-\wt+\subrphase) \pol_{\submode}.
\end{equation}
where $\veck=\veck_{ijn}$. It is worth noticing that when the total number of wave vectors $N_{\veck}$ becomes very large, both uniform spherical and cubic sampling effectively sample all the wave vectors $\veck$ in $\veck$-space. Therefore, in the limit of large sampling number \disp{N_{\veck} \rightarrow \infty}, the two sampling methods become equivalent\footnote{
The relation $ \rho_{\veck} = 1/\Delta^3k = N_{\veck}/V_{\veck} = V/(2\pi)^3$ implies that the limit of large sampling number (i.e.\ $N_{\veck} \rightarrow \infty$) is equivalent to the limit of unbounded space (i.e.\ $V \rightarrow \infty$). At this limit, the volume element becomes differential (denoted as $d^3k$) and is free from any specific shape associated with the space boundary. Therefore, all sampling methods for the allowed wave vectors $\veck$ become equivalent, and the summation form approaches the integral form. This is consistent with the fact that no volume factor $V$ is involved in the integral form of the RED vacuum field in unbounded space, as shown in \Eq{Boyer:vacE}.}, and \Eq{dnsty} can be used for both sampling methods to calculate the volume factor $V$ in Eqs.~(\ref{vacEcubic}) and (\ref{vacEspherical}),
\begin{equation}\label{vol}
V = (2\pi)^3\rho_{\veck} = (2\pi)^3\frac{N_{\veck}}{V_\veck},
\end{equation}
where
\begin{equation}\label{kvol} 
V_{\veck} = \frac{4\pi}{3}\left(\frac{\wo+\Delta/2}{c}\right)^3 - \frac{4\pi}{3}\left(\frac{\wo-\Delta/2}{c}\right)^3.
\end{equation}

	In a computer simulation, the summation indices in \Eq{vacEspherical} can be rewritten as
\begin{equation}\label{vacEnested}
\vect{E}_{vac} =\sum_{\lambda=1}^{2} \sum_{i=1}^{N_{\kappa}} \sum_{j=1}^{N_{\vartheta}} \sum_{n=1}^{N_{\varphi}} \sqrt{\frac{\hbar\omega}{\eps V}} \cos(\kdr-\wt+\subrphase) \pol_{\submode},
\end{equation}
where the multiple sums indicate a numerical nested loop, and the wave vector $\veck=\veck_{ijn}$ is chosen according to the uniform spherical sampling method. In practice, $N_{\vartheta}$ and $N_{\varphi}$ need to be sufficiently large so that the wave vector $\veck$ at a fixed frequency may be sampled isotropically. Also, a large $N_{\kappa}$ is required for representative samplings in frequency. As a result, $N_{\veck} = N_{\kappa}N_{\vartheta}N_{\varphi}$ has to be very large to achieve convergence when using uniform spherical sampling.
Therefore, sampling $\veck$ for many angles at each frequency makes RED simulation computationally intensive. To improve the efficiency of our simulation, we sample $\veck$ for only one random angle $(\theta_i, \phi_i)$ at each frequency (i.e.\ $N_{\kappa}=N_{\omega}$, $N_{\vartheta}= 1$, and $N_{\varphi}=1$). Namely, $N_{\veck}=N_{\omega}$, and for $i = 1\ldots N_{\omega}$,
\begin{equation}\label{mslct:layer1}
\veck_{i}
=
\begin{pmatrix}
k^{(x)}_{i} \\ k^{(y)}_{i} \\ k^{(z)}_{i} 
\end{pmatrix}
=
\begin{pmatrix}
k_{i}\sin{\theta_i}\cos{\phi_i} \\
k_{i}\sin{\theta_i}\sin{\phi_i} \\
k_{i}\cos{\theta_i}
\end{pmatrix},
\end{equation}
where 
\begin{equation}\label{mslct:layer2}
\left\{ 
\begin{array}{l}
k_{i} = (3\kappa_{i})^{1/3} \\
\theta_{i} = \cos^{-1}(\vartheta_{i}) \\
\phi_{i} = \varphi_{i},
\end{array}
\right.
\end{equation}
and
\begin{equation}\label{mslct:layer3}
\left\{ 
\begin{array}{l}
\kappa_{i} = (\wo-\Delta/2)^3/3c^3 + (i-1)\dkapa \\
\vartheta_{i} = R_{i}^{(1)} \\
\varphi_{i} = R_{i}^{(2)}.
\end{array}
\right.
\end{equation}
The stepsize $\dkapa$ is specified in \Eq{dka}, $\vartheta = R^{(1)}$ is a random number uniformly distributed in $[-1,1]$, and $\varphi = R^{(2)}$ is another random number uniformly distributed in $[0,2\pi]$. As the number of sampled frequencies $N_{\omega}$ becomes sufficiently large, the random angles $(\theta, \phi)$ will be uniformly distributed on a unit sphere, and the angular distribution of the wave vectors $\veck$ is isotropic. 

	In the limit \disp{\frac{\Delta}{\wo} \ll 1}, the above sampling method (described by Eqs.~(\ref{mslct:layer1}), (\ref{mslct:layer2}), and (\ref{mslct:layer3})) and the uniform spherical sampling method both approach a uniform sampling on a spherical surface at the radius \disp{r_k = \frac{\wo}{c}}. In this limit, the addition of the condition in \Eq{resonshell} leads to the possible choices of the frequency range $\Delta$,
\begin{equation}\label{reson}
\Gamma\wo \ll \frac{\Delta}{\wo} \ll 1.
\end{equation}
Within this range (\Eq{reson}), the expression for the RED vacuum field in \Eq{vacE} becomes	
\begin{equation}\label{vacEsingle}
\vect{E}_{vac} =\sum_{\lambda=1}^{2} \sum_{i=1}^{N_{\omega}} \sqrt{\frac{\hbar\omega}{\eps V}} \cos(\kdr-\wt+\subrphase) \pol_{\submode}.
\end{equation} 
where $\veck=\veck_{i}$. For large $N_{\veck}=N_{\omega}$, the volume factor $V$ is calculated using \Eq{vol}.

\begin{figure}[t]
\centering
\scalebox{0.45}{\includegraphics{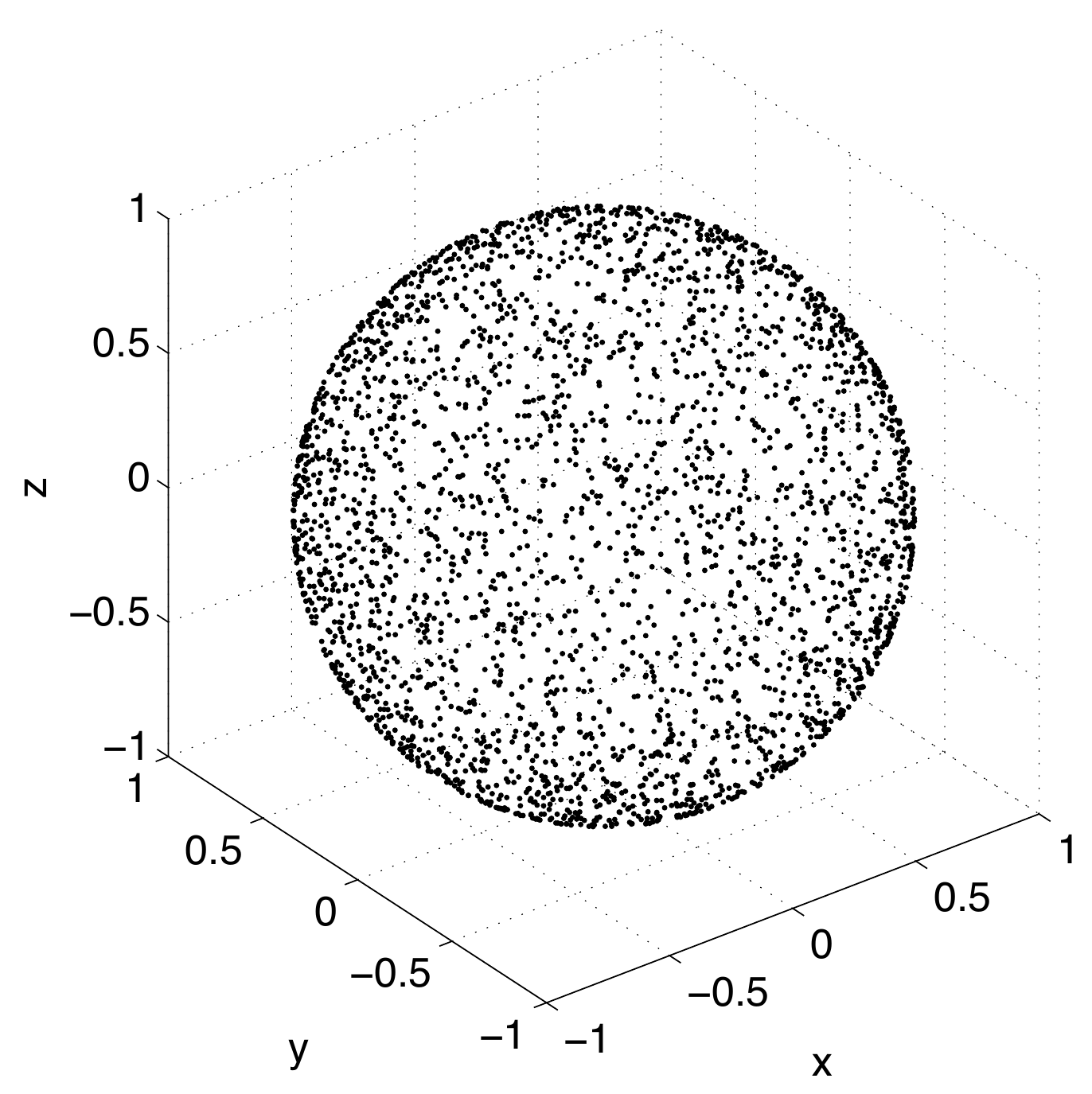}}
\caption{The isotropic distribution of the polarization field vectors $\pol_{_{\veck,1}}$ (number of sampled frequencies number $N_{\omega}$ = 3k). The endpoints of the polarization field vectors, $\pol_{_{\veck,1}}$, are plotted for a random sampling of modes $({\veck,1})$ according to the methods described in Eqs.~(\ref{mslct:layer1}), (\ref{mslct:layer2}), and (\ref{mslct:layer3}).}
\label{fig:Edist} 
\end{figure}

	Finally, for a complete specification of the simulated RED vacuum field, \Eq{vacEsingle}, the polarizations $\pol_{\submode}$ need to be chosen. From \Eq{vol},  we notice that large $N_{\veck}$ gives large $V$. Since for large $V$ the RED vacuum field is not affected by the space boundary, there is no preferential polarization direction. Therefore, the polarizations are isotropically distributed. The construction for isotropically distributed polarizations is discussed in detail in \App{app:isotropy}. Here we give the result that satisfies the property of isotropy and the properties of polarization (described by Eqs.~(\ref{pol:cond_1}) and (\ref{pol:cond_2})),
\begin{equation}\label{pol}
\begin{split}
&\pol_{\submodeup} = 
\begin{pmatrix}
\varepsilon_{\submodeup}^{(x)} \\ \varepsilon_{\submodeup}^{(y)} \\ \varepsilon_{\submodeup}^{(z)}
\end{pmatrix}
= 
\begin{pmatrix}
\cos{\theta_i}\cos{\phi_i}\cos{\chi_i}-\sin{\phi_i}\sin{\chi_i} \\
\cos{\theta_i}\sin{\phi_i}\cos{\chi_i}+\cos{\phi_i}\sin{\chi_i} \\
-\sin{\theta_i}\cos{\chi_i}
\end{pmatrix}       
\\	
\\			      
&\pol_{\submodedown} =  
\begin{pmatrix}
\varepsilon_{\submodedown}^{(x)} \\ \varepsilon_{\submodedown}^{(y)} \\ \varepsilon_{\submodedown}^{(z)}
\end{pmatrix}
=
\begin{pmatrix}
-\cos{\theta_i}\cos{\phi_i}\sin{\chi_i}-\sin{\phi_i}\cos{\chi_i} \\
-\cos{\theta_i}\sin{\phi_i}\sin{\chi_i}+\cos{\phi_i}\cos{\chi_i} \\
\sin{\theta_i}\sin{\chi_i}
\end{pmatrix}
,
\end{split}
\end{equation}
where $\chi$ is a random number uniformly distributed in $[0,2\pi]$. With the wave vectors $\veck$ (described by Eqs.~(\ref{mslct:layer1}), (\ref{mslct:layer2}), and (\ref{mslct:layer3})) and the polarizations $\pol_{\submode}$ (described by \Eq{pol}), the endpoints of the sampled RED vacuum field vector are plotted on a unit sphere as shown in \Fig{fig:Edist}, which illustrates the isotropy of the distribution. 
	
	In summary, the RED vacuum field, \Eq{vacEsingle}, can be specified by a set of four numbers $(\kappa_i, \vartheta_i, \varphi_i, \chi_i)$, chosen according to Eqs.~(\ref{mslct:layer1}), (\ref{mslct:layer2}), (\ref{mslct:layer3}), and \Eq{pol}. The only assumption used in determining these numbers is \Eq{reson}, which is equivalent to the sharp resonance condition, \Eq{sharpreson}, used in Boyer's analysis. 

\subsection{Simplified Equation of Motion}

	An RED harmonic oscillator is a charged classical harmonic oscillator immersed in the RED vacuum field. The complete equation of motion is
\begin{equation}\label{EOM}
m\ddot{x}= -m\wo^2x + m\Gamma\dddot{x} + e\left[ E_{vac}^{(x)}(x,t) + \left( \vect{v} \times \vect{B}_{vac}(x,t) \right)^{(x)} \right],
\end{equation}
where $m$ is the mass, $e$ is the charge, and $\wo$ is the natural frequency. The radiation damping parameter is \disp{\Gamma \equiv \frac{2e^2}{3mc^3}\frac{1}{4\pi\eps}}. Only the motion in the $x$-direction is considered in this paper, so the position of the particle is \disp{\vect{r} = (x, 0, 0)}, and the velocity of the particle is \disp{\vect{v} = (\dot{x}, 0, 0)}. Also, the $x$-component of the vector $\vect{v} \times \vect{B}_{vac}$ is denoted as $\left( \vect{v} \times \vect{B}_{vac} \right)^{(x)}$.

	Three terms are involved in the equation of motion --- the spring force, radiation damping, and the RED vacuum field, 
\begin{align}
&F_{spring} = -m\wo^2x, \label{spring} \\
&F_{damp} = m\Gamma\dddot{x}, \label{damp}\\
&F_{vac} = e\left[ E_{vac}^{(x)} + \left( \vect{v} \times \vect{B}_{vac} \right)^{(x)} \right]. \label{vac}
\end{align}
In unbounded space, the RED vacuum field is given by \Eq{Boyer:vacE}. In a bounded space of volume $V$, the RED vacuum field, including both the electric part $\vect{E}_{vac}$ as in \Eq{vacE} and the magnetic part $\vect{B}_{vac}$, is
\begin{equation}\label{vacEB}
\begin{split}
&\vect{E}_{vac}(\vect{r},t) = \sum_{\mode} \sqrt{\frac{\hbar\omega}{\eps V}} \cos(\kdr-\wt+\subrphase) \pol_{\submode}, \\
&\vect{B}_{vac}(\vect{r},t) = \frac{1}{c} \sum_{\mode} \sqrt{\frac{\hbar\omega}{\eps V}} \cos(\kdr-\wt+\subrphase) ( {\hat{\veck} \times \pol_{\submode}} ),
\end{split}
\end{equation}
where $\omega = c|\veck|$, and $\subrphase$ is the random phase uniformly distributed in $[0, 2\pi]$. Throughout this section the RED vacuum field as described by \Eq{vacEB} will be used.

	Before a simulation of the RED harmonic oscillator is performed, we may increase the efficiency of our computation by making some approximations to the equation of motion, \Eq{EOM}. Namely, the terms $F_{damp}$, $F_{vac}$ can be approximated by using the dipole approximation $\kdr \ll 1$ (\Eq{dipoleapprox}), and the sharp resonance condition $\Gamma\wo \ll 1$ (\Eq{sharpreson}).
		
	Denoting the order of magnitude for length, time, and electric field as $\xs$, $\ts$, and $\Es$ respectively, the order of magnitude of the two terms of $F_{vac}$ in \Eq{vac} are evaluated\footnote{
The order of magnitude for dimension $A$ is denoted as $[A]$.},
\begin{equation}\label{order:original}
\begin{split}
&[eE_{vac}^{(x)}] = e\Es, \\
&[e( \vect{v} \times \vect{B}_{vac} )^{(x)}] =\left(\frac{\xs}{c\ts}\right)e\Es.
\end{split}
\end{equation}	
The sharp resonance condition \Eq{sharpreson} makes the harmonic oscillator responsive to only frequencies that are close to $\wo$, thus the time scale of the particle's motion is \disp{\ts \simeq \frac{1}{\wo}}. The dimensionless parameter \disp{\frac{\xs}{c\ts}} in \Eq{order:original} can thus be evaluated, \disp{\frac{\xs}{c\ts} \simeq k_0\xs}, to yield
\begin{equation}\label{order:freq}
\begin{split}
&[eE_{vac}^{(x)}] = e\Es, \\
&[e( \vect{v} \times \vect{B}_{vac} )^{(x)}] = (k_0\xs) e\Es.
\end{split}
\end{equation}
Furthermore, the dipole approximation from \Eq{dipoleapprox} implies 
\begin{equation}
k_0\xs \ll 1,
\end{equation}
which can be used to approximate $F_{vac}$ with its dominating term,
\begin{equation}\label{vac:1st-order}
F_{vac} \simeq e\Es.
\end{equation}
The dipole approximation also removes the spatial dependence in the RED vacuum field,
\begin{gather}\label{vacE:1st-order}
\begin{split}
\vect{E}_{vac} = & \sum_{\mode} \pol_{\submode} \sqrt{\frac{\hbar\omega}{\eps V}} \left[ \left( 1- \frac{(\kdr)^2}{2!}+\ldots \right)\cos(\wt-\subrphase) \right. \\
		        	& \left. +\left( \kdr- \frac{(\kdr)^3}{3!}+\ldots \right)\sin(\wt-\subrphase)  \right] \\
	     \simeq & \sum_{\mode} \pol_{\submode} \sqrt{\frac{\hbar\omega}{\eps V}}\cos(\wt-\subrphase) \\
	     \simeq & \sum_{\mode} \sqrt{\frac{\hbar\omega}{\eps V}} \frac{1}{2} \left(a_{\submode}e^{-i\wt} + a^{\ast}_{\submode}e^{i\wt}\right) \pol_{\submode},
\end{split}\\
a_{\submode} \equiv e^{i\subrphase} \nonumber.
\end{gather}

	Through the approximations, Eqs.~(\ref{vac:1st-order}) and (\ref{vacE:1st-order}), the equation of motion \Eq{EOM} becomes
\begin{equation}\label{EOM:vac-1st}
m\ddot{x}= -m\wo^2x + m\Gamma\dddot{x} + eE_{vac}^{(x)}(t).
\end{equation}
The radiation damping $m\Gamma\dddot{x}$ in \Eq{EOM:vac-1st} is well known to give runaway solutions. Boyer's approach is to find the solution through Fourier transformation \cite{Boyer:RED&QM}. Following this approach, the steady-state solution for \Eq{EOM:vac-1st} may be obtained, 
\begin{equation}\label{sol:1st-order}
x(t) = \frac{e}{m}\sum_{\mode} \sqrt{\frac{\hbar\omega}{\eps V}} \frac{1}{2} \left( \frac{a_{\submode}}{C_{\submode}}e^{-i\wt} + \frac{a^{\ast}_{\submode}}{C^{\ast}_{\submode}}e^{i\wt}\right) \varepsilon^{(x)}_{\submode}, \\
\end{equation}
where $C_{\submode} \equiv \left(-\omega^2 + \wo^2 \right) - i\Gamma\omega^3$. 
Although \Eq{sol:1st-order} could be evaluated numerically, this would not allow for straightforward generalization to physical systems where the equation of motion is modified, which is one of the purposes of this work. Hence, we follow the approach in \cite{Landau&Lifshitz, Jackson, Poisson} to avoid runaway solutions. According to classical electrodynamics, the equation of motion for an electron with radiation damping is
\begin{equation}\label{eq1}
m\ddot{x} = F_{ext} + m\gamma\dddot{x},
\end{equation}
which can be approximated with
\begin{equation}\label{eq2}
m\ddot{x} \simeq F_{ext}, 
\end{equation}
using the assumption $m\gamma\dddot{x} \ll F_{ext}$. Such an assumption is necessary for a point-particle description of the electron to be valid \cite{Landau&Lifshitz, Poisson}. With the reduced equation (\Eq{eq2}), the radiation damping $m\gamma\dddot{x}$ may be estimated as
\begin{equation}
m\gamma\dddot{x} \simeq \gamma \dot{F}_{ext},
\end{equation}
and then iterated back to the original equation (\Eq{eq1}),
\begin{equation}\label{eq3}
m\ddot{x} \simeq F_{ext} + \gamma\dot{F}_{ext}.
\end{equation}
This approximate equation of motion is free of runaway solutions.
Applying \Eq{eq3} to the RED harmonic oscillator, \Eq{EOM:vac-1st}, we obtain
\begin{equation}\label{EOM:damp}
m\ddot{x} \simeq -m\wo^2x -m\Gamma\wo^2\dot{x} + eE_{vac}^{(x)} + e\Gamma\dot{E}_{vac}^{(x)}.
\end{equation}
The order of magnitude of each force term is
\begin{equation}
\begin{split}
F_{spring}  &\longrightarrow [m\wo^2x] = m\wo^2\xs , \\
F_{vac}      &\longrightarrow [eE_{vac}^{(x)}] = e\Es, \\
F_{damp}  &\longrightarrow
	\left\{
	\begin{split}
	&[m\Gamma\wo^2\dot{x}] = (\Gamma\wo)m\wo^2\xs \\
	&[e\Gamma\dot{E}_{vac}^{(x)}] = (\Gamma\wo) e\Es. \\
	\end{split}
	\right. \\
\end{split}
\end{equation}

	To evaluate the magnitude of the terms in $F_{damp}$, a random walk model may be used. For a fixed time $t=t_0$, a particular value of $E_{vac}^{(x)}(t_0)$ and $x(t_0)$ are found in Eqs.~(\ref{vacE:1st-order}) and (\ref{sol:1st-order}), and the order of magnitude for $E_{vac}^{(x)}(t_0)$ and $x(t_0)$ are equal to $\Es$ and $\xs$. The mathematical form of the complex $\tilde{E}_{vac}^{(x)}(t_0)$ and $\tilde{x}(t_0)$ is analogous to a two-dimensional random walk on the complex plane with random variable $\Theta_{\{\mode\}}$, where ${\{\mode\}}$ denotes a set of modes $(\mode)$. Thus, if averaged over ${\{\mode\}}$, the order of magnitude for $E_{vac}^{(x)}(t_0)$ and $x(t_0)$ may be estimated by the root-mean-squared distance of $\tilde{E}_{vac}^{(x)}(t_0)$ and $\tilde{x}(t_0)$. In a two-dimensional random walk model \cite{random walk}, the root-mean-squared distance $D_{rms}$ is given by
\begin{equation}
D_{rms} = \sqrt{N_{s}} \cdot \Delta s,
\end{equation}
where $N_{s}$ is the number of steps taken, and $\Delta s$ is a typical stepsize; for $D_{rms}^{(E)}$, \disp{\Delta s =\frac{1}{2} \sqrt{\frac{\hbar\wo}{\eps V}}}, and for $D_{rms}^{(x)}$, \disp{\Delta s =\frac{1}{2} \left( \frac{e}{m\Gamma\wo^3}\sqrt{\frac{\hbar\wo}{\eps V}} \right)}. Hence, the order of magnitude, $\Es$ and $\xs$, may be estimated as\footnote{
Using $V = (2\pi)^3N_{\omega}/V_{\veck}$ and $V_{\veck} \simeq 4\pi\wo^2\left( \Gamma\wo^2 \right)/c^3$ (Eqs.~(\ref{vol}) and (\ref{kvol})), the value of $\xs$ can be estimated as $\xs \simeq \sqrt{3/\pi} \sqrt{\hbar/2m\wo}$, which is consistent with Boyer's calculation for the standard deviation of position in \Eq{length}.}
\begin{equation}\label{order:xsEs}
\begin{split}
&\Es \simeq D_{rms}^{(E)} = \sqrt{2N_{\omega}} \cdot \frac{1}{2}\sqrt{\frac{\hbar\wo}{\eps V}}, \\
&\xs \simeq D_{rms}^{(x)} = \sqrt{2N_{\omega}} \cdot \frac{1}{2}\left( \frac{e}{m\Gamma\wo^3}\sqrt{\frac{\hbar\wo}{\eps V}} \right).
\end{split}
\end{equation}
The order of magnitude for $F_{damp}$ is evaluated accordingly,
\begin{equation}
\begin{split}
&[m\Gamma\wo^2\dot{x}] \simeq e \sqrt{\frac{N_{\omega}}{2}}\cdot\sqrt{\frac{\hbar\wo}{\eps V}} \\
&[e\Gamma\dot{E}_{vac}^{(x)}] \simeq (\Gamma\wo) e \sqrt{\frac{N_{\omega}}{2}}\cdot\sqrt{\frac{\hbar\wo}{\eps V}}. \\
\end{split}
\end{equation}
Again, using the sharp resonance condition $\Gamma\wo \ll 1$ (\Eq{sharpreson}), $F_{damp}$ is approximated with its dominating term, 
\begin{equation}
F_{damp} \simeq e \sqrt{\frac{N_{\omega}}{2}}\cdot\sqrt{\frac{\hbar\wo}{\eps V}},
\end{equation}
and the equation of motion, \Eq{EOM:damp}, is further simplified to\footnote{
Both Boyer's equation of motion, \Eq{EOM:Boyer}, and the equation of motion used in our work, \Eq{EOM:sim}, give the same result on the standard deviation (\Eq{length}) and the moments (\Eq{moment}).}
\begin{equation}\label{EOM:sim}
m\ddot{x} \simeq -m\wo^2x -m\Gamma\wo^2\dot{x} + eE_{vac}^{(x)}(t).
\end{equation}

	As an additional note, given the estimation of $\Es$ and $\xs$ in \Eq{order:xsEs}, the three force terms in \Eq{EOM:sim} have the following relation, 
\begin{equation}
[m\wo^2x] \gg [m\Gamma\wo^2\dot{x}] \simeq [eE_{vac}^{(x)}].
\end{equation}
The implication is that the spring force $F_{spring}$ dominates, while the RED vacuum field $F_{vac}$ and the radiation damping $F_{damp}$ are perturbations. The driving force $F_{vac}$ and the damping force $F_{damp}$ will establish a balance, keeping the amplitude of the motion of the RED harmonic oscillator in the vicinity of $\xs$.

	Finally, after the main equations for simulation are completely defined (Eqs.~(\ref{vacE:1st-order}) and (\ref{EOM:sim})), the range of the integration time $\tau_{int}$ (i.e.\ how long the simulation runs) needs to be chosen. Upon careful inspection, two important time scales can be identified from the analytical solution of \Eq{EOM:sim},
\begin{equation}\label{sol:sim}
\begin{split}
x(t) =&\; e^{-\Gamma\wo^2t/2}(Ae^{i\omega_Rt}+A^{\ast}e^{-i\omega_Rt}) \\
	&+ \frac{e}{m}\sum_{\mode} \sqrt{\frac{\hbar\omega}{\eps V}} \frac{1}{2} \left( \frac{a_{\submode}}{\mathcal{C}_{\submode}}e^{-i\wt} + \frac{a^{\ast}_{\submode}}{\mathcal{C}^{\ast}_{\submode}}e^{i\wt}\right) \varepsilon^{(x)}_{\submode},
\end{split}
\end{equation}
where $A$ is a coefficient determined by the initial conditions, and
\begin{gather}
 \omega_R \equiv \wo\sqrt{1-\left( \frac{\Gamma\wo}{2} \right)^2}, \\
 \mathcal{C}_{\submode} \equiv \left( -\omega^2 + \wo^2 \right) - i\left(\Gamma\wo^2 \right)\omega, \\
 a_{\submode} \equiv e^{i\subrphase}.
\end{gather} 
The first term in \Eq{sol:sim} is the transient part of the solution. The characteristic time for the transient motion is  
\begin{equation}\label{trans}
\tau_{tran} = \frac{2}{\Gamma\wo^2}.
\end{equation}
The position should be evaluated for $\tau_{trans} \ll \tau_{int}$, if we are only interested in the steady-state motion. The second term in \Eq{sol:sim} is the steady-state part of the solution. Because the steady-state solution is a finite discrete sum of periodic functions, it would have a non-physical repetition time $\tau_{rep}$. The choice of $\tau_{int}$ should satisfy $\tau_{int} \le \tau_{rep}$ to avoid repetitive solutions. A detailed discussion about $\tau_{rep}$ can be found in \App{app:repetition}. Here we give the choice of $\tau_{int}$,
\begin{equation}\label{rep}
\tau_{int} = \frac{2\pi}{\dw}, 
\end{equation}
where $\dw$ is the gap between adjacent frequencies. The frequency gap $\dw$ can be estimated using Eqs.~(\ref{dka}), (\ref{mslct:layer2}), and (\ref{mslct:layer3}),
\begin{equation}
\dw \simeq \frac{c(3\kappa_0)^{1/3}}{3}\frac{\dkapa}{\kappa_0},
\end{equation}
where \disp{\kappa_0 = \frac{1}{3} \left( \frac{\wo}{c} \right)^3}.

	To summarize, \Eq{EOM:sim} is the equation of motion that will be used for simulating the RED harmonic oscillator. The RED vacuum field in \Eq{EOM:sim} is simulated by \Eq{vacE:1st-order}, and the specifications of its modes ($\mode$), polarizations $\pol_{\submode}$, and other relevant variables can be found in \Sec{sec:RED vacuum field setup}. The simplification from \Eq{EOM} to \Eq{EOM:sim} relies on only two conditions --- namely, the dipole approximation \Eq{dipoleapprox} and the sharp resonance condition \Eq{sharpreson}, which are also the only two conditions used in Boyer's analysis \cite{Boyer:RED&QM}. The parameters $e,m$, and $\wo$ should be chosen to satisfy these two conditions. Lastly, the integration time $\tau_{int}$ of the simulation is to be chosen within the range $\tau_{trans} \ll \tau_{int} \le \tau_{rep}$, where $\tau_{tran}$ and $\tau_{rep}$ are given in Eqs.~(\ref{trans}) and (\ref{rep}) respectively.

\section{Results}\label{sec:Results}

	In \Sec{sec:RED probability}, it was shown that the RED harmonic oscillator probability distribution is Gaussian, contrasting to the double-peak probability distribution of the classical harmonic oscillator. We set out to investigate how the dynamics of the RED harmonic oscillator can give rise to such a Gaussian distribution. In this section, the result of the simulation for an RED harmonic oscillator is presented, and the relation between the trajectory and the probability distribution is discussed.  
	
	In constructing the probability distribution from the particle trajectory, two sampling methods are used, namely sequential sampling and ensemble sampling. In a sequential sampling, the position or velocity is recorded in a time sequence from one trajectory, while in an ensemble sampling, the same is recored only at the end of the simulation from an ensemble of trajectories. The recorded positions or velocities are collected in histogram and then converted to a probability distribution, which is compared to the analytical result, \Eq{prob}. Whereas the sequential sampling illustrates the relationship between the buildup of the probability distribution and the dynamics of the trajectory, the ensemble sampling is convenient for statistical interpretation. In addition, the ensemble sampling is suitable for parallel computing, which improves the computation efficiency.

\begin{figure}[t]
\centering
\scalebox{0.5}{\includegraphics{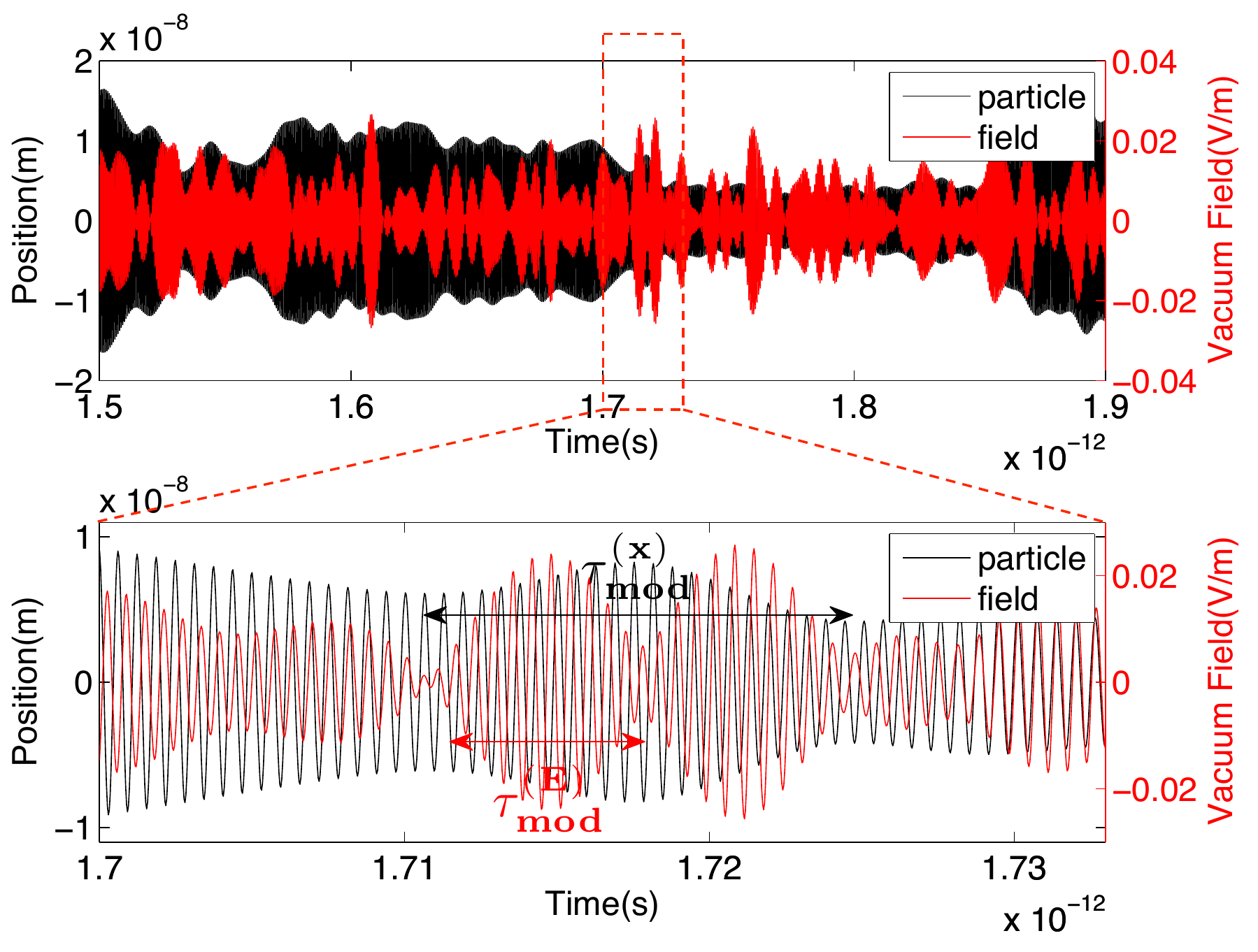}}
\caption{A comparison between the field evolution and the particle trajectory. Top: RED vacuum field (red solid line) is compared to the trajectory of the RED harmonic oscillator (black solid line). Bottom: A magnified section of the trajectory shows that there is not a fixed phase or amplitude relation between the particle trajectory (black solid line) and the instantaneous driving field (red solid line). The modulation time for the field (red solid line) is also shown to be longer than that for the motion of the harmonic oscillator (black solid line).}
\label{fig:xE}
\end{figure}

\begin{figure}[t]
\centering
\scalebox{0.5}{\includegraphics{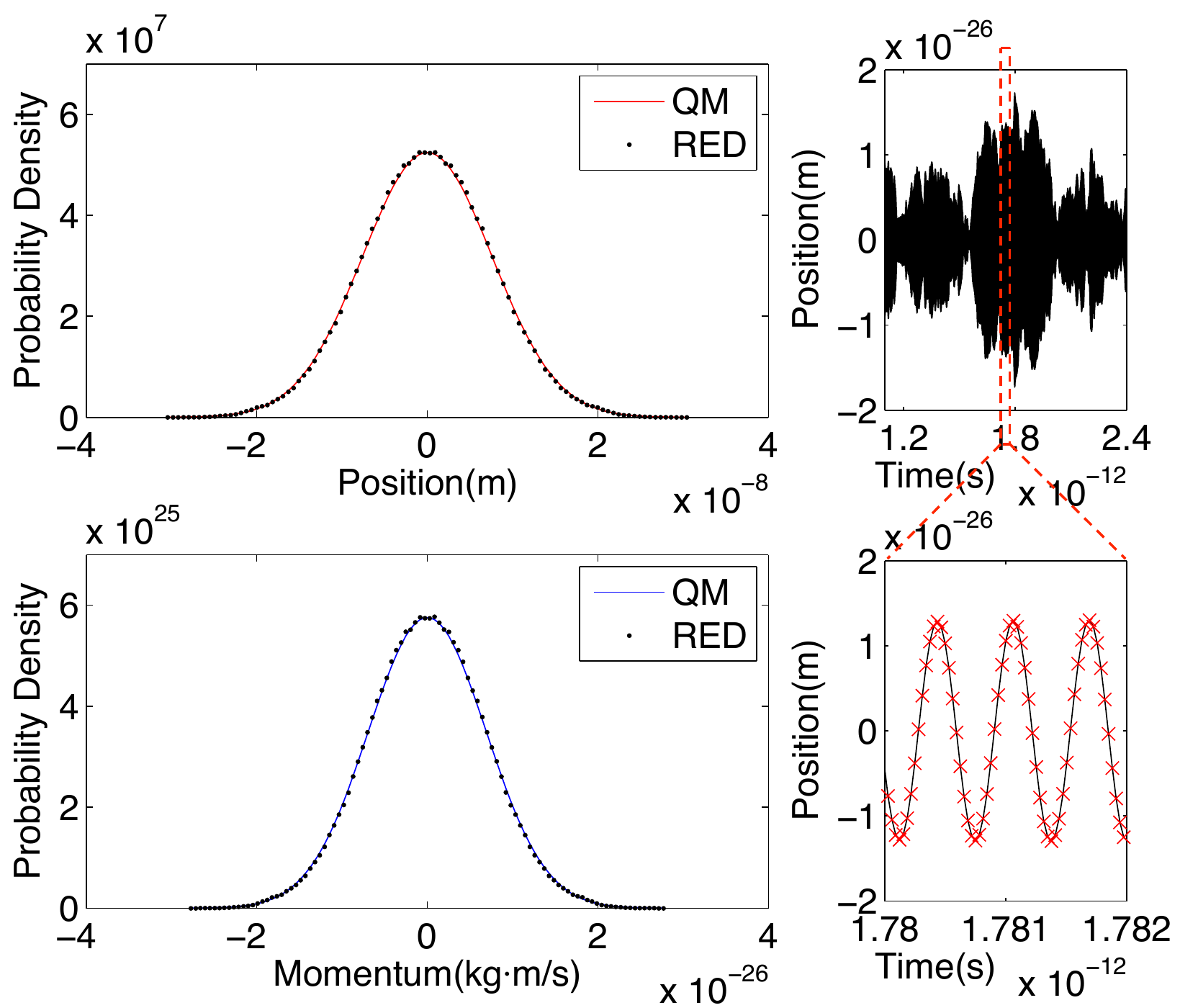}}
\caption{The RED probability distribution constructed from a sequential sampling (number of sampled frequencies $N_{\omega}$ = 20k). Left: The position and momentum probability distributions for the RED harmonic oscillator (black dot) and the ground state quantum harmonic oscillator (red \& blue solid line) are compared. Right: A illustration of the sequential sampling shows how positions are recorded at a regular time sequence (red cross). Note that at small time scale the oscillation amplitude is constant, but at large time scale it modulates.}
\label{fig:time}
\end{figure}

\begin{figure}[t]
\centering
\scalebox{0.5}{\includegraphics{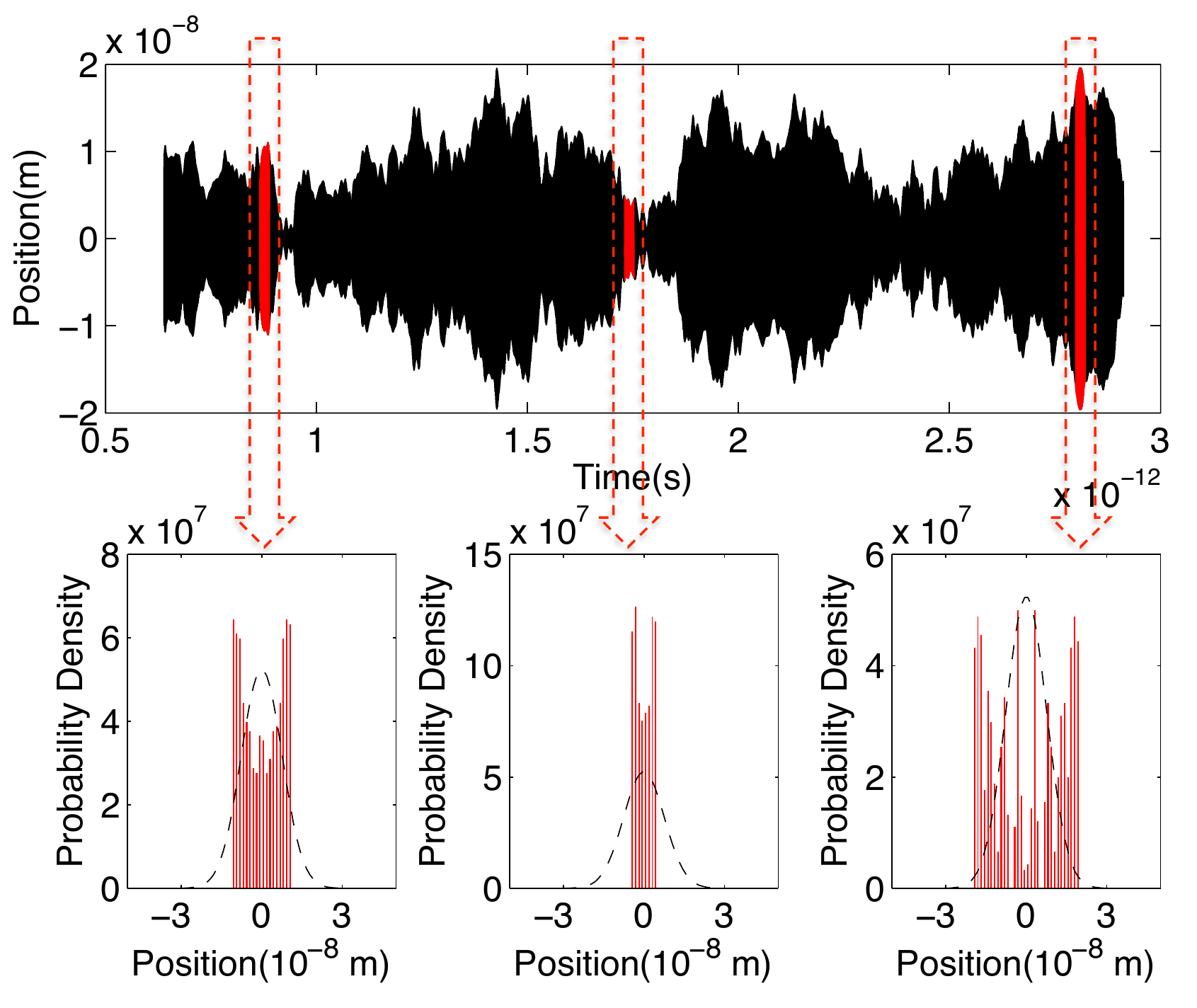}}
\caption{Contributions of different oscillation amplitudes in the final probability distribution. Top: A steady-state trajectory of the RED harmonic oscillator (black) and its several sections (red) are shown. A section is limited to the duration of a modulation time. The oscillation amplitude changes significantly outside a modulation time, so different sections have different oscillation amplitude. Bottom: The probability distributions of each trajectory section. Since the oscillation amplitude is approximately constant in each section, the corresponding probability distribution (red bar) is close to the classical double-peak distribution. These probability distributions contribute to different areas of the final probability distribution (black dash line) that is constructed from the full steady-state trajectory.}
\label{fig:trajhist}
\end{figure}

\begin{figure}[t]
\centering
\scalebox{0.45}{\includegraphics{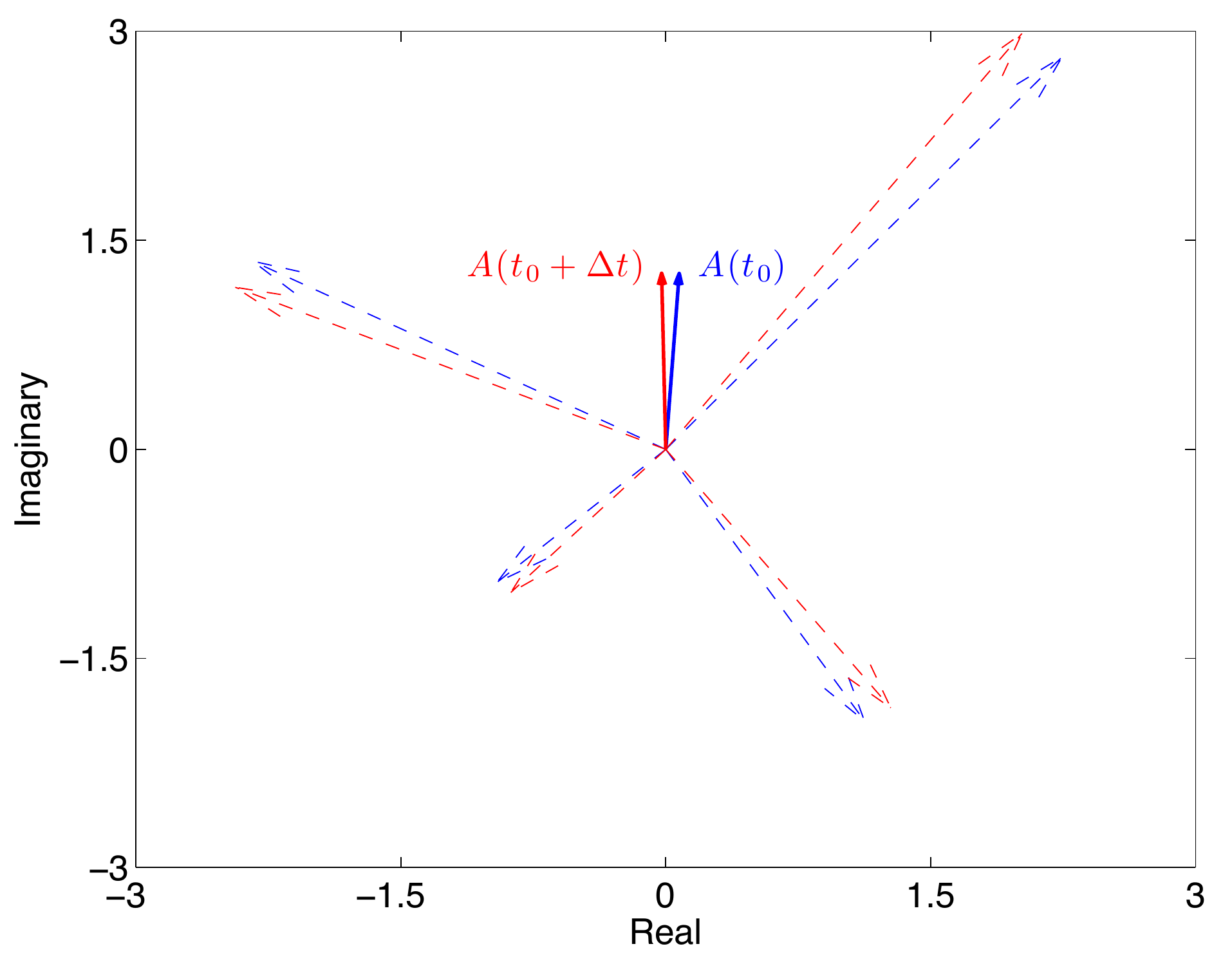}}
\caption{A schematic illustration of the oscillation amplitude as a sum of different frequency components in the complex plane. At a particular time $t=t_0$, an oscillation amplitude $A(t_0)$ (blue solid arrow) is formed by a group of frequency components (blue dash arrow), which rotate in the complex plane at different rate. After a time $\Delta t \ll \tau_{coh}$, the angles of the frequency components (red dash arrow) change only a little. Therefore, the oscillation amplitude $A(t_0+\Delta t)$ (red solid arrow) does not change much within the coherence time.}
\label{fig:cohtime}
\end{figure}
		
\begin{figure*}
\centering
\scalebox{0.7}{\includegraphics{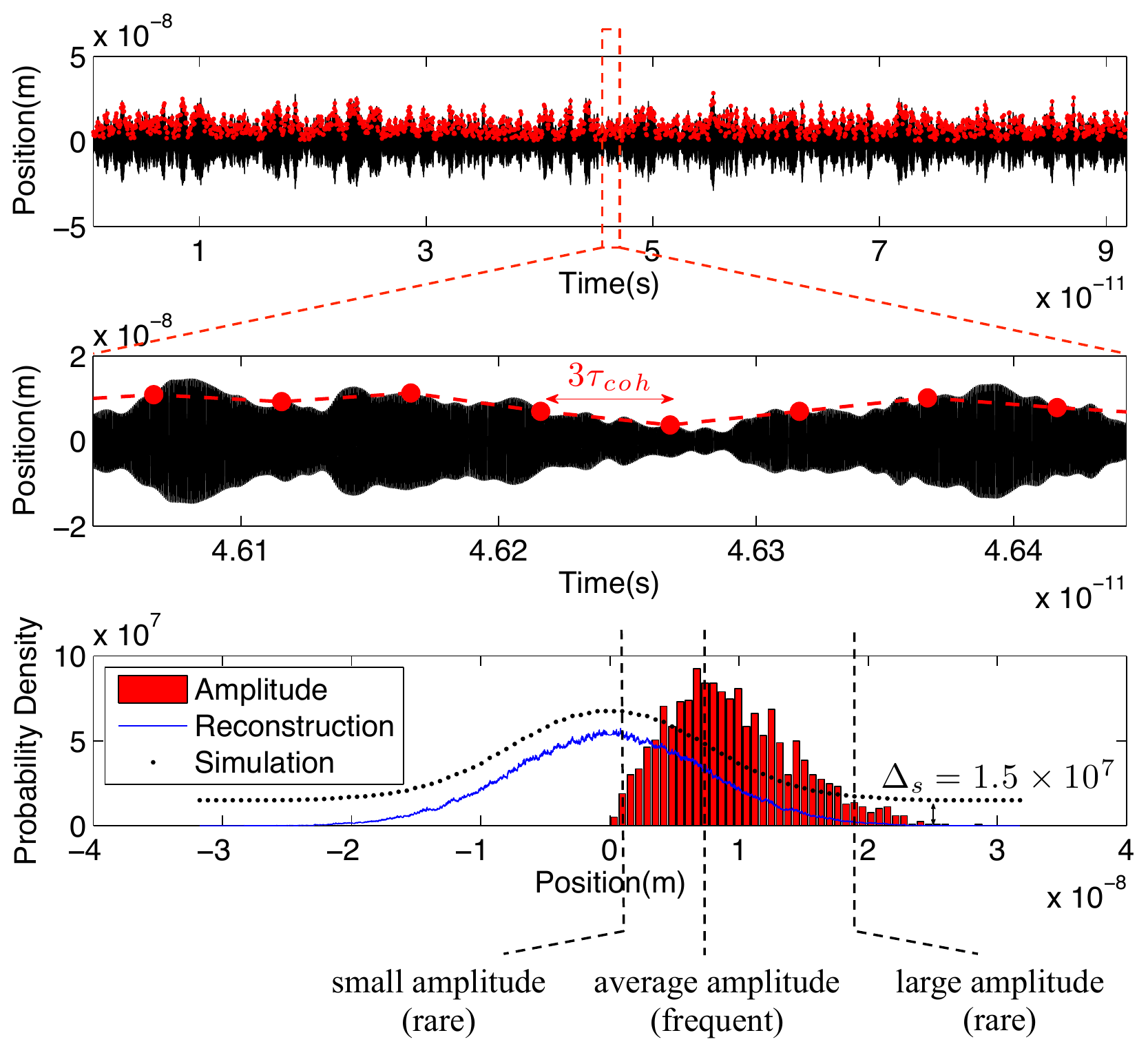}}
\caption{Sampled oscillation amplitude and the reconstructed position probability distribution. Top: The oscillation amplitudes (red dot) are sampled from the full steady-state particle trajectory (black) with a sampling time-step larger than one coherence time $\tau_{coh}$. Middle: The sampling time-step for the oscillation amplitudes (red dot) is approximately $3\tau_{coh}$. Bottom: The probability distributions of the position (black dot) and the oscillation amplitude (red bar) are constructed from the simulated data. Notice that the most frequently occurring oscillation amplitude corresponds to the half-maximum width of the position distribution (black dot), which is offset by $\Delta_{s}=1.5\times 10^{7}$. The reconstructed position probability distribution (blue solid line) is given by \disp{P(x)=\sum_{A} P_{A}(x) = \int \! P_{A}(x)f(A)dA}, where $P_{A}(x)$ is the classical double-peak probability distribution associated with an oscillation amplitude $A$, and $f(A)$ is the probability distribution of the oscillation amplitude. This procedure demonstrates that the Gaussian RED probability distribution can be thought of as a sum of classical double-peak distributions with a specific amplitude distribution, answering the question illustrated in \Fig{fig:CMREDintro}.}
\label{fig:amp}
\end{figure*}

\begin{figure}[t]
\centering
\scalebox{0.5}{\includegraphics{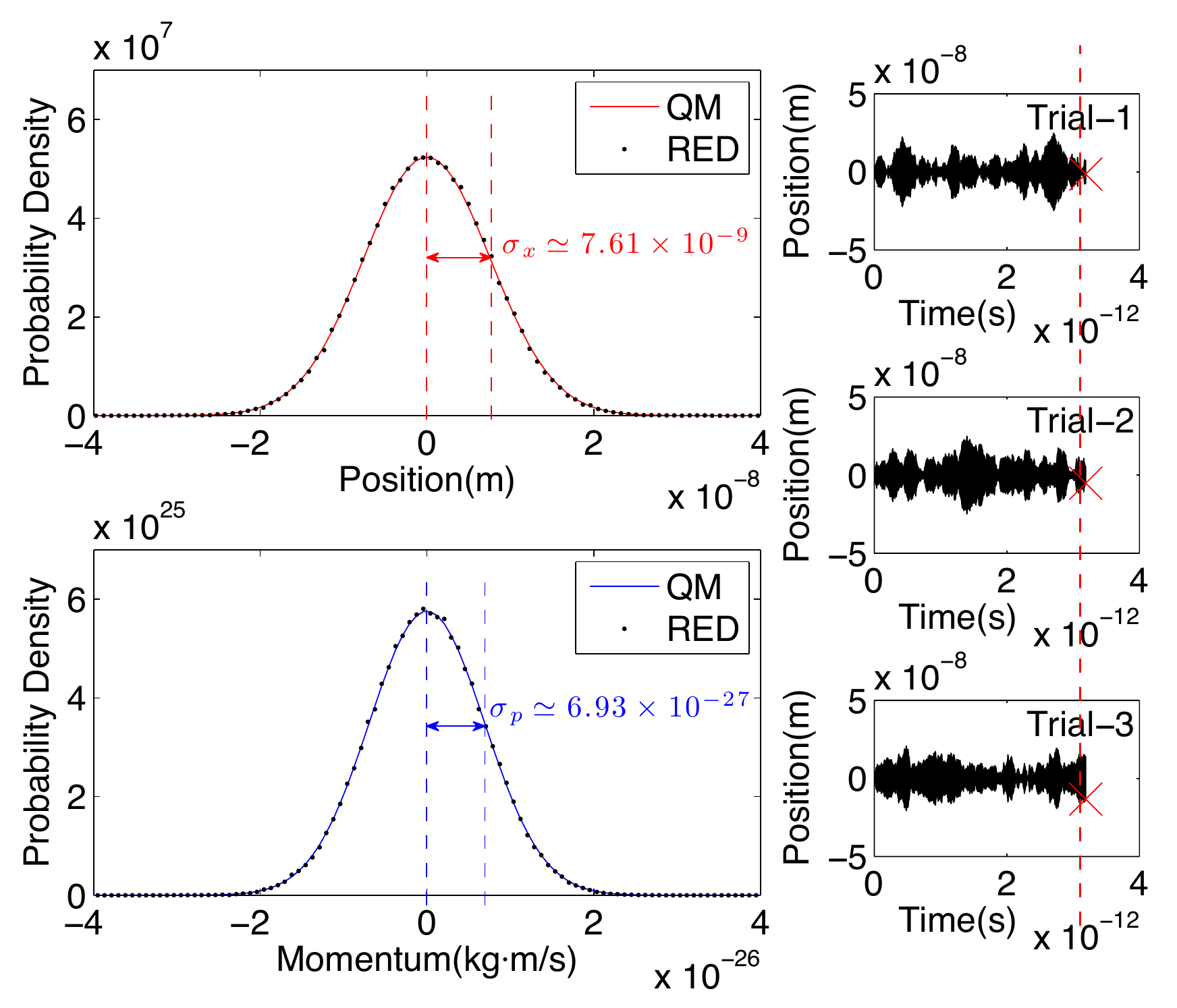}}
\caption{The RED probability distribution constructed from an ensemble sampling (number of particles $N_{p}$ = 200k, number of sampled frequencies $N_{\omega}$ = 2k) in contrast to the sequential sampling (\Fig{fig:time}). Left: The position and momentum probability distributions are shown for the RED harmonic oscillator (black dot) and the ground state quantum harmonic oscillator (red \& blue solid line). As the ensemble sampling corresponds to the procedure of random phase averaging (Eqs.~(\ref{length}) and (\ref{momentum})), the widths of the two RED distributions satisfy Heisenberg's minimum uncertainty relation, $\sigma_x \sigma_p = \hbar/2$. Right: A illustration of the ensemble sampling shows how positions (red cross) are recorded from an ensemble of trajectories (black).}
\label{fig:ensm}
\end{figure}

\begin{figure}[t]
\centering
\scalebox{0.5}{\includegraphics{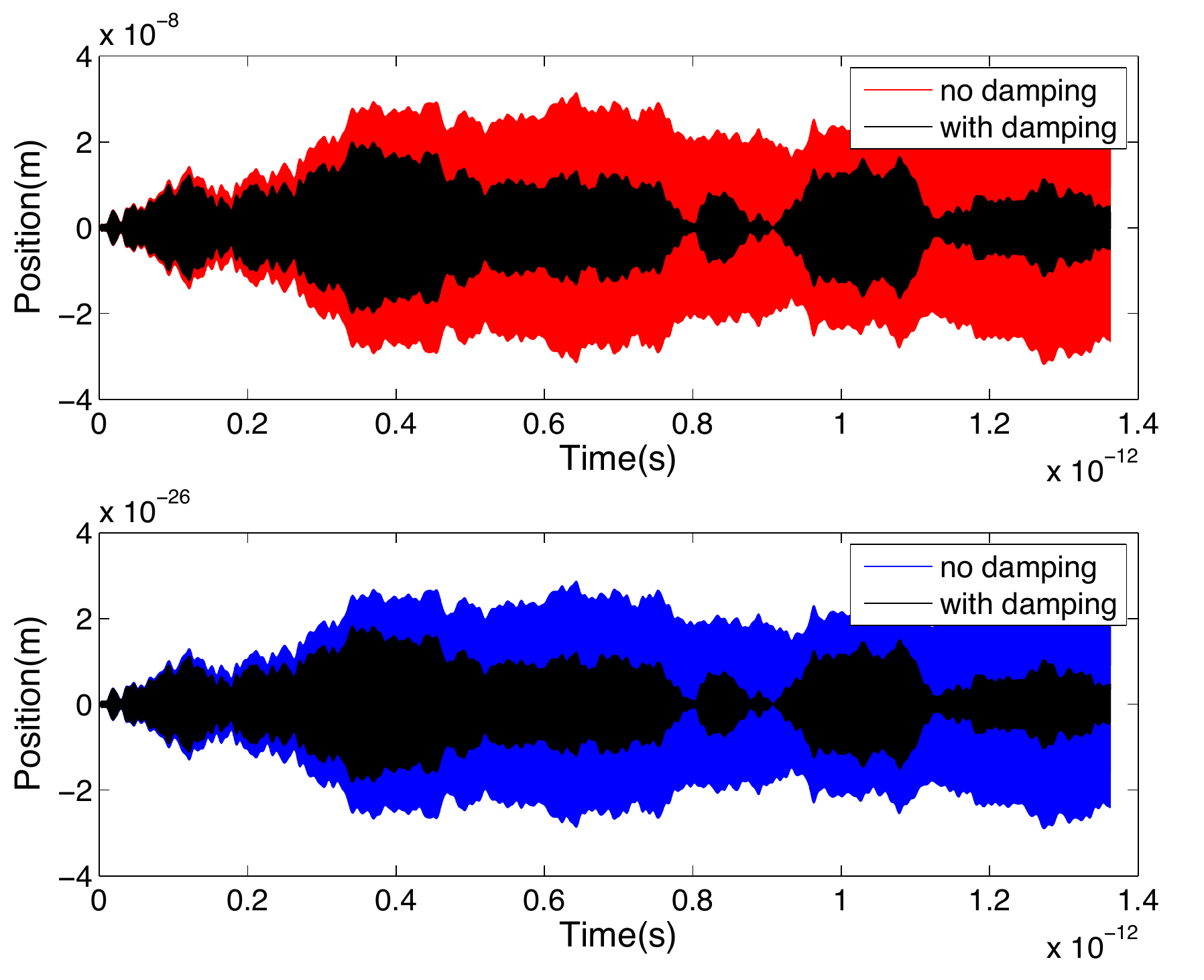}}
\caption{Radiation damping and Heisenberg's minimum uncertainty relation. Under the balance between vacuum field driving and the radiation damping, the trajectory of the RED harmonic oscillator (black solid line) satisfies the Heisenberg minimum uncertainty relation. When the radiation damping is turned off in the simulation (red \& blue solid line), the minimum uncertainty relation no longer holds, although the range of the particle's motion is still constrained by the harmonic potential.}
\label{fig:diverge}
\end{figure}

\subsection{Sequential Sampling and the Relation to its Probability Distribution}

	By solving \Eq{EOM:sim} numerically, the steady-state trajectory for the RED harmonic oscillator is obtained as shown in \Fig{fig:xE}. For a comparison, the RED vacuum field is also included. The differential equation (\Eq{EOM:sim}) is solved using the adaptive 5th order Cash-Karp Runge-Kutta method \cite{recipe}, and the integration stepsize is set as one twentieth of the natural period, \disp{\frac{1}{20}\left(\frac{2\pi}{\wo}\right)}. The charge $e$, mass $m$, and natural frequency $\wo$ are chosen as $q_e$, $10^{-4}m_e$, and $10^{16}$ (rad/s) respectively, where $q_e$ is the electron charge and $m_e$ is the electron mass. The frequency range $\Delta$ of the RED vacuum field is chosen to be much broader than the resonance width, $220\times \Gamma\wo^2$. The choice of $10^{-4}m_e$ is made to bring the modulation time and the natural period of the harmonic oscillator closer to each other. In other words, the differential equation (\Eq{EOM:sim}) covers time scales at two extremes, and the choice of mass $10^{-4}m_e$ brings these scales closer to each other while keeping the integration time manageable without losing the physical characteristics of the problem. Some features of the trajectory, shown in \Fig{fig:xE}, are worth mentioning. First, there is not a fixed phase or amplitude relation between the particle trajectory and the instantaneous driving field. Second, the rate of amplitude modulation in the particle trajectory is slower than that in the driving field. While these results may seem a little surprising at first glance, the steady-state solution in \Eq{sol:sim} provides some insights into both of these behaviors. 	
		
	To understand the features of the trajectory, we study the temporal response of the harmonic oscillator as revealed in the Green formulation of the steady-state solution \cite{Marion},
\begin{equation}\label{sol:green}
x(t) = \frac{e}{m\omega_R}\int_{-\infty}^t E_{vac}^{(x)}(t')e^{-\Gamma\wo^2(t-t')/2}\sin{(\omega_R(t-t'))}dt' .
\end{equation}
The solution indicates that the driving field \disp{E_{vac}^{(x)}(t')} at any given time $t'$ has an effect on the particle for a finite period of time, \disp{\frac{1}{\Gamma\wo^2}}. In other words, the particle trajectory $x(t)$ at a particular time $t$ is affected by the driving field \disp{E_{vac}^{(x)}(t')} from all previous moments $t' \le t$. This explains why the particle trajectory does not seem to have a fixed phase and amplitude relation with the instantaneous driving field. Another implication of \Eq{sol:green} is that it always takes a finite period of time, \disp{\frac{1}{\Gamma\wo^2}}, for the particle to dissipate the energy gained from the driving field. Thus, even if the field already changes its amplitude, it would take a while before the particle can respond. This is why the amplitude modulation in the particle trajectory is slower compared to that in the driving field\footnote{
However, in case of slow field modulation (i.e.\ field bandwidth shorter than resonance width of the harmonic oscillator) the modulation time of the field and the particle are the same.}.

	Using a sequential sampling method, a probability distribution is constructed from the simulated trajectory of a RED harmonic oscillator as shown in \Fig{fig:time}. The RED probability distribution turns out to be a Gaussian distribution, which is the same as the ground state probability distribution of a quantum harmonic oscillator. To understand the relation between the trajectory and the Gaussian probability distribution, we investigate the dynamics of the RED harmonic oscillator at two time scales. First, at small time scale, the particle oscillates as a classical harmonic oscillator. The oscillation amplitude is constant, and the period is $T = 2\pi/\wo$. Such an oscillation makes a classical double-peak probability distribution. Second, at large time scale, the oscillation amplitude modulates. As a result, different parts of the trajectory have double-peak probability distributions associated with different oscillation amplitudes, which add to make the final probability distribution a Gaussian distribution as shown in \Fig{fig:trajhist}.

	To further understand how amplitude modulation turns the final probability distribution into a Gaussian distribution, we again inspect the steady-state solution in \Eq{sol:sim},
\begin{equation}\label{sol:steady}
x(t) = \frac{e}{m}\sum_{\mode} \sqrt{\frac{\hbar\omega}{\eps V}} \frac{1}{2} \left( \frac{a_{\submode}}{\mathcal{C}_{\submode}}e^{-i\wt} + c.c. \right) \varepsilon^{(x)}_{\submode}.
\end{equation}
The solution can be rewritten in terms of the complex position $\tilde{x}(t)$, 
\begin{equation}
\begin{split}
x(t) &= Re\left( \tilde{x}(t) \right), \\
\tilde{x}(t) &= \frac{e}{m}\sum_{\mode} \sqrt{\frac{\hbar\omega}{\eps V}} \frac{a_{\submode}}{\mathcal{C}_{\submode}}e^{-i\wt} \varepsilon^{(x)}_{\submode}.
\end{split}
\end{equation}
Since the frequencies are sampled symmetrically around $\wo$, the amplitude modulation is made evident as the complex position $\tilde{x}(t)$ is factorized into a modulation term $A(t)$ and an oscillation term $e^{-i\wo t}$,
\begin{equation}
\begin{split}
\tilde{x}(t) &= A(t)e^{-i\wo t}, \\
A(t) &= \sum_{\mode} \left( \varepsilon^{(x)}_{\submode} \frac{e}{m}\sqrt{\frac{\hbar\omega}{\eps V}} \frac{a_{\submode}}{\mathcal{C}_{\submode}} \right) e^{-i(\omega-\wo)t}.
\end{split}
\end{equation}
The amplitude $A(t)$ is a sum of many complex components, which rotate in the complex plane at different rates, $\omega-\wo$. At any given time, the amplitude $A(t)$ is determined by the configuration of these components, as shown in \Fig{fig:cohtime}. As time elapses, the configuration of the components evolves and the amplitude $A(t)$ may change dramatically. However, when the elapsed time $\Delta t$ is much smaller than the coherence time $\tau_{coh}$, which is given by the shortest rotating period, the change of amplitude $A(t)$ is negligible,
\begin{gather}
A(t+\Delta t) \simeq A(t) \quad \mbox{for} \quad \Delta t \ll \tau_{coh}, \label{amp}\\
\tau_{coh} = \left[ \frac{2\pi}{|\omega-\wo|} \right]_{min} \label{coh}.
\end{gather}
Therefore, since an oscillation amplitude $A(t)$ is approximately constant within one coherence time $\tau_{coh}$\footnote{
The coherence time $\tau_{coh}$ defined in this work has the same meaning as the temporal width of the first-order correlation function. We can take the autocorrelation of the simulated trajectory and show that its temporal width is the same as the coherence time computed in this work.}, the amplitude modulation as previously discussed can be characterized by the coherence time $\tau_{coh}$. In other words, the coherence time $\tau_{coh}$ is the modulation time. 	

	Now, if we take a representative sampling of the oscillation amplitudes (i.e.\ each sampled amplitude is several coherence times $\tau_{coh}$ apart), we obtain an amplitude distribution as shown in \Fig{fig:amp}. From the amplitude distribution, it is clear that the occurrence of very large or very small amplitudes is rare. This is because it requires all the components of the amplitude $A(t)$ to be either completely aligned or completely misaligned to obtain the extreme values of $A(t)$. For most of the time, the configuration is in partial alignment, and the amplitude $A(t)$ is of the average size as calculated previously (i.e.\ $\xs$). Using the amplitude distribution in \Fig{fig:amp}, a Gaussian distribution can be reconstructed by adding up all the classical double-peak distributions associated with these sampled oscillation amplitudes. This reconstructed Gaussian probability distribution resembles the probability distribution obtained directly from the simulated data as shown in \Fig{fig:amp}. It is worth pointing out that it is impossible to reconstruct such a Gaussian distribution with any amplitude distribution other than the one given by the particle trajectory (\Fig{fig:amp}). This result shows that the Gaussian probability distribution of the RED harmonic oscillator is a consequence of its intrinsic amplitude modulation.

\subsection{Ensemble Sampling and Heisenberg Uncertainty Relation}

	In many RED analyses \cite{Boyer:RED, Boyer:diamagnetism, Boyer:unrah, Boyer:casimir, Boyer:RED&QM, Boyer:anharmonic}, the procedure of random phase averaging is an essential tool in obtaining the statistical properties of a physical system. To compare the RED numerical results with the analyses, it is advantageous to use ensemble sampling. In ensemble sampling, particles in each trajectory of the ensemble are prepared with identical initial conditions, but the RED vacuum fields differ in the initial random phases $\subrphase$. Here, different random initial phases $\subrphase$ represent the physical realization of random phase averaging. At the end of the simulation, physical quantities such as position and momentum are recorded from an ensemble of trajectories as shown in \Fig{fig:ensm}. In the case of RED harmonic oscillator, the recorded positions and momentum are converted into probability distributions, and the Heisenberg minimum uncertainty relation is found to be satisfied. Boyer suggested that the minimum uncertainty relation is established by a delicate balance between the gain of energy from RED vacuum field driving and the loss of energy through radiation damping \cite{Boyer:RED}. As we turn off the radiation damping in the simulation (\Fig{fig:diverge}), the minimum uncertainty relation no long holds, although the range of the particle's motion is still constrained by the harmonic potential. This result supports the balancing mechanism proposed by Boyer.
	
	Unlike sequential sampling, ensemble sampling has the advantage that the recorded data are fully uncorrelated. As a result, the integration time does not need to be much longer than one coherence time $\tau_{coh}$ after the particle reaches its steady-state. However, this does not mean that the computation time will be shorter. In fact, because only one data point is recorded from each trajectory, a simulation with ensemble sampling usually takes much longer than with sequential sampling. For example, a typical RED simulation with sequential sampling (i.e.\ number of sampled frequencies $N_{\omega}=20k$) takes 2.3 hours to finish, but a typical RED simulation with ensemble sampling (i.e.\ number of particles $N_p=200k$, number of sampled frequencies $N_{\omega}=0.5k$) takes 61 hours, which is 27 times longer.
		
\begin{figure}[t]
\centering
\scalebox{0.45}{\includegraphics{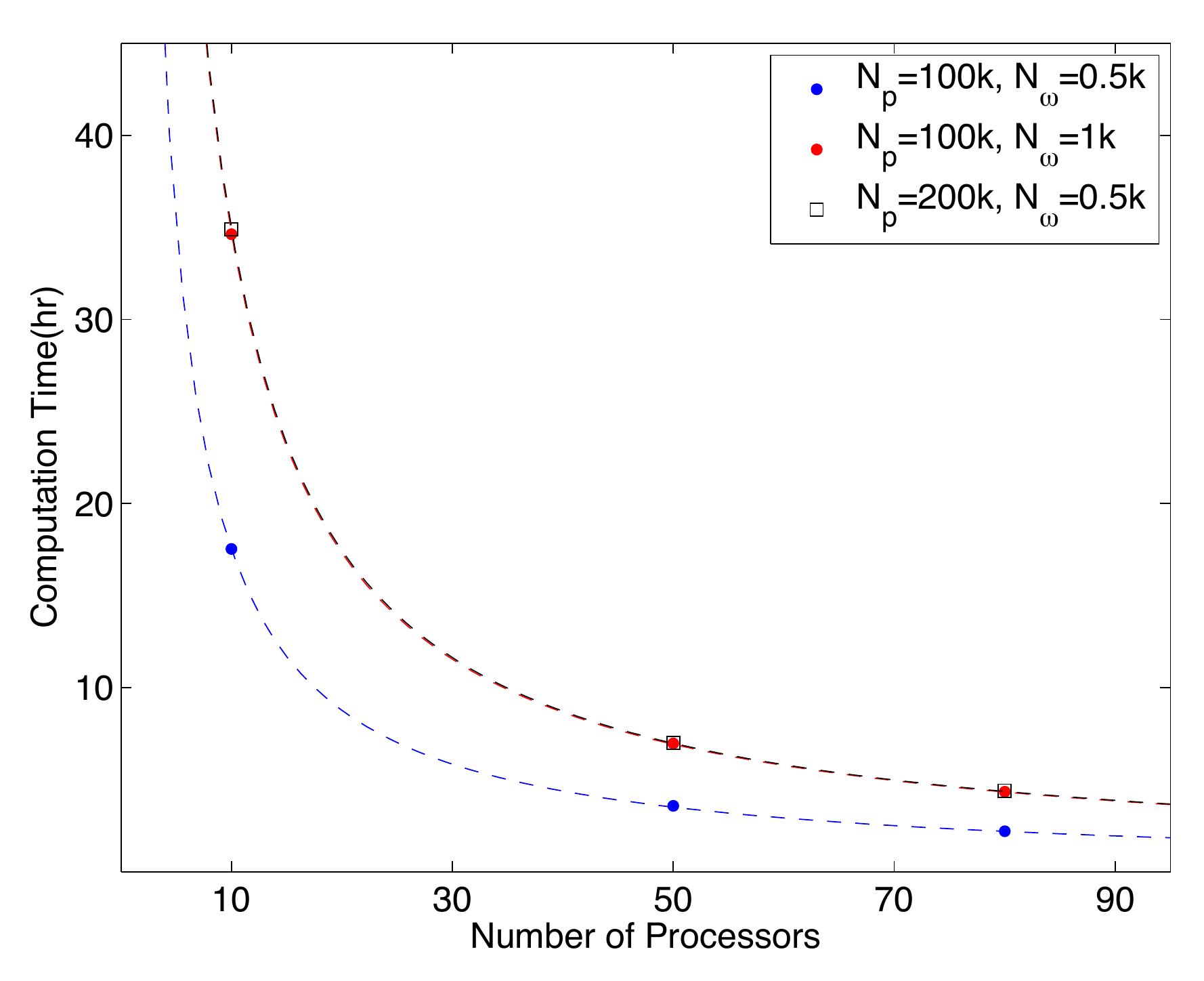}}
\caption{The inverse relation between the computation time and the number of processors. The data (black open square, blue \& red dot) follow an inverse relation $y = C/x$ (black, blue \& red dash line). The parameter $C$ is determined by the product of $N_p$ (number of particles) and $N_{\omega}$ (number of sampled frequencies).}
\label{fig:computation}
\end{figure}

\begin{figure}[t]
\centering
\scalebox{0.45}{\includegraphics{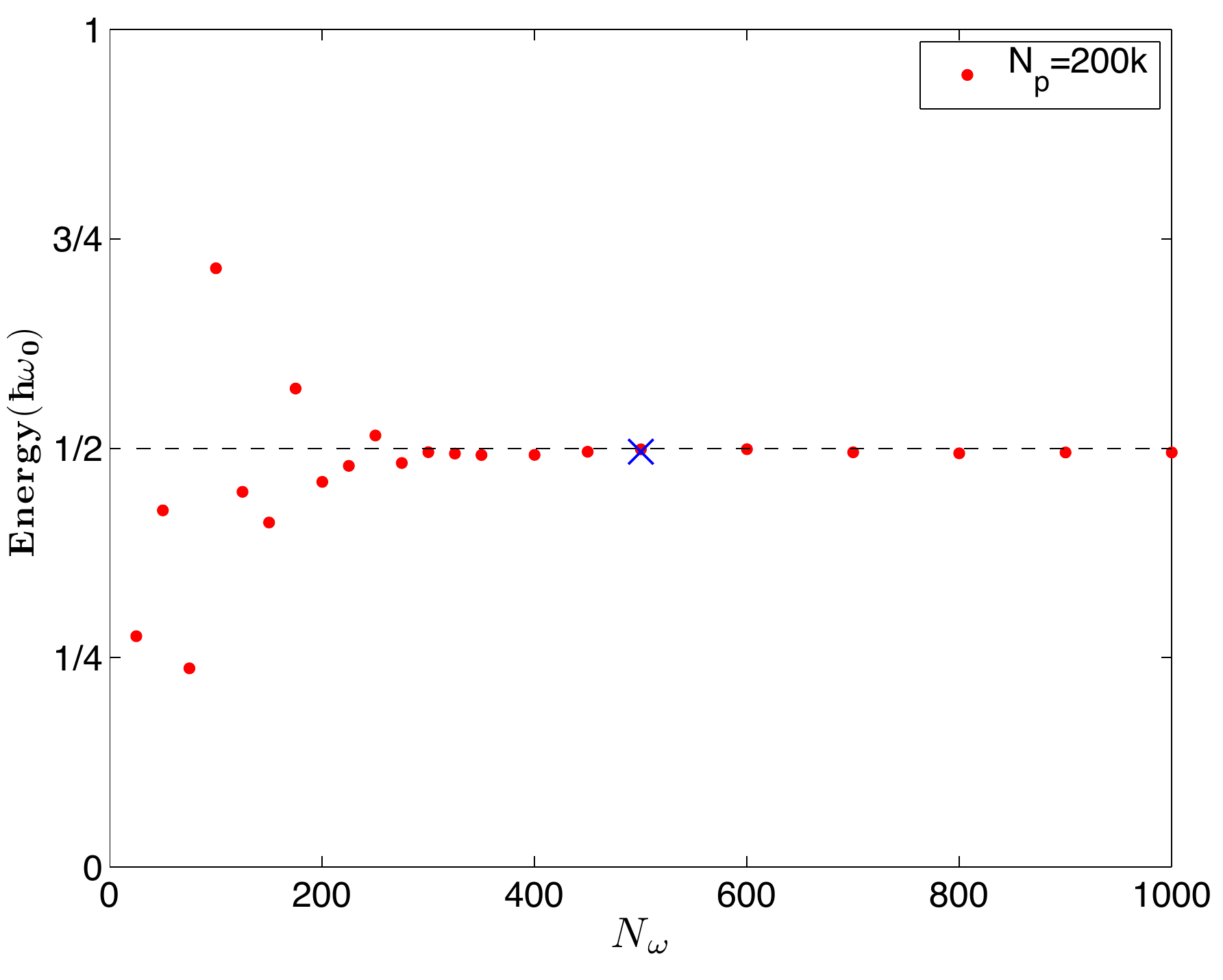}}
\caption{The ensemble-averaged energy of the RED harmonic oscillator and its convergence as a function $N_{\omega}$. The number of particles is $N_{p}$ = 200k, and the sampled frequency range is $\Delta = 220 \times \Gamma\wo^2$. The energy converges as the number of sampled frequencies $N_{\omega}$ increases. At $N_{\omega}$ = 0.5k (blue cross), the deviation from the ground state energy of quantum harmonic oscillator is 1\%.}
\label{fig:convergence}
\end{figure}

	A resolution to this problem is to incorporate parallel computing in the ensemble sampling approach. The parallelization of the RED simulation is straightforward for ensemble sampling, since each trajectory are independent except for the random initial phases $\subrphase$. To reduce the amount of interprocessor communication and computation overhead, each processor is assigned an equal amount of work. A typical simulation with ensemble sampling approach (i.e.\ number of particles $N_{p}$ = 200k, number of sampled frequencies $N_{\omega}$ = 0.5k) takes 61 hours to run on a single processor. After parallelization, it takes only 4.5 hours to run on 80 processors, which is a factor of 14 speed-up\footnote{
The parallelization of the RED simulation program is developed and benchmarked with assistance from University of Nebraksa Holland Computing Center. The program is written in Fortran and parallelized using Message Passing Interface (MPI) \cite{recipe,MPI}. The compiler used in this work is from the GNU Compiler Collection (GCC) ``\texttt{openmpi-1.4.3/gcc-4.4.1}''. However, with the commercial compiler ``\texttt{openmpi-1.4.4/intel-11.1}'', we have found that the computation time can be further reduced by a factor of 6.}, and the computation time scales inversely with the number of processors as shown in \Fig{fig:computation}. In addition, the parallelized code is advantageous for testing the numerical convergence of the simulation. In \Fig{fig:convergence}, the convergence of the ensemble-averaged energy as a function of sampled frequency number is shown.

\section{Summary \& Conclusions}\label{sec:Summary}

\begin{figure*}
\centering
\scalebox{0.55}{\includegraphics{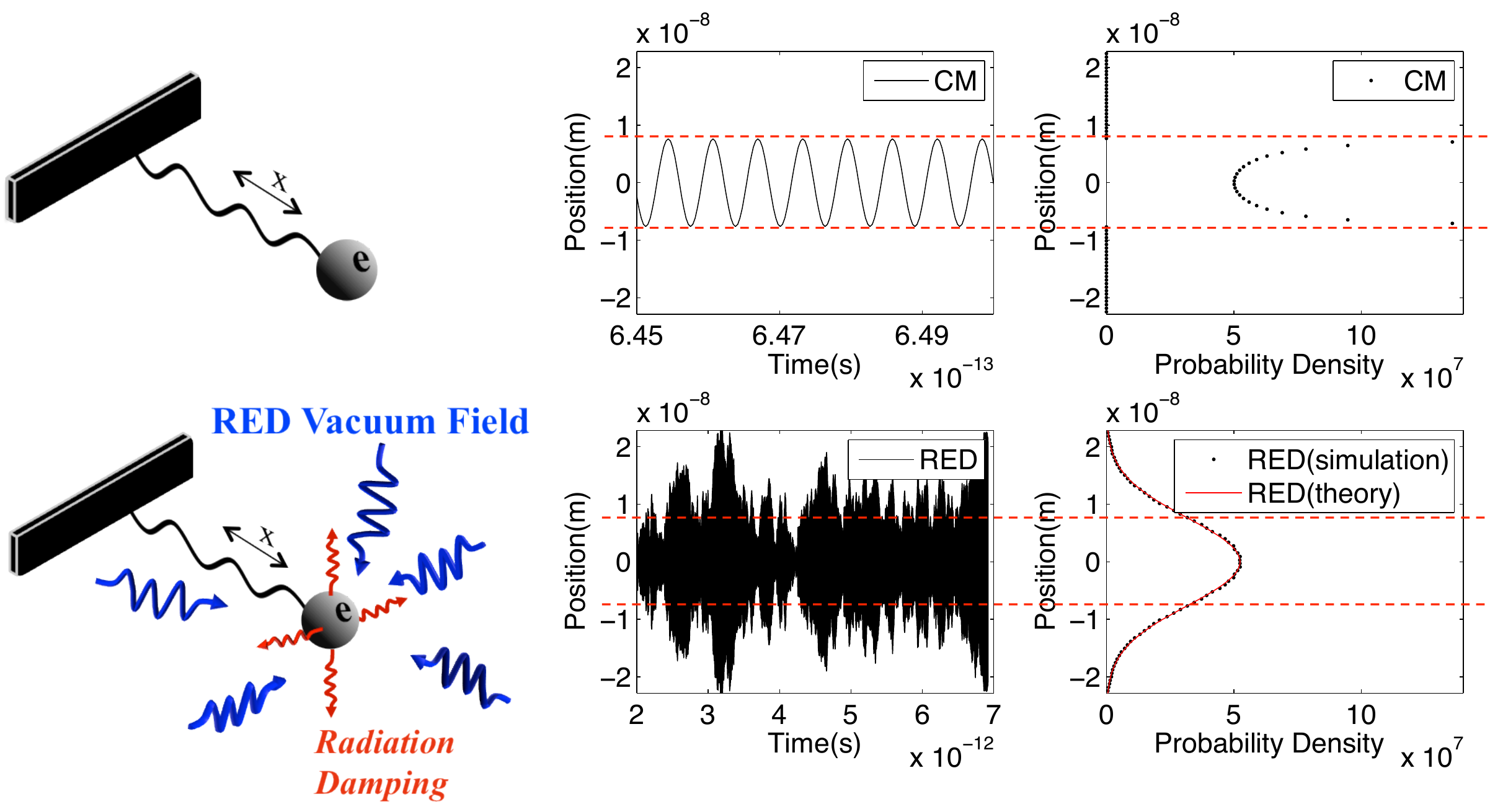}}
\caption{A comparison between the classical and the RED harmonic oscillators. Top: A free classic harmonic oscillator that is initially displaced from equilibrium demonstrates a motion of constant amplitude, which results in double-peak probability distribution. Bottom: Compared to \Fig{fig:CMREDintro}, it is clear now that the RED harmonic oscillator undergoes an oscillatory motion with modulating oscillation amplitude. The oscillation amplitude modulates at the time scale of coherence time $\tau_{coh}$ and is responsible for the resulting Gaussian probability distribution.}
\label{fig:CMREDfinal}
\end{figure*}

	The analytical probability distribution of an RED harmonic oscillator is obtained in \Sec{sec:RED probability}. The numerical method for the RED simulation is documented in \Sec{sec:RED simulation}. Agreement is found between the simulation and the analytical results. Namely, the probability distributions constructed from the simulated data follows the same Gaussian distributions as the ground state quantum harmonic oscillator (Figs.~\ref{fig:time}, and \ref{fig:ensm}). As a consequence, the Heisenberg minimum uncertainty relation is satisfied by the widths of the RED position and momentum distributions. In the absence of radiation damping, this minimum uncertain relation no longer holds (\Fig{fig:diverge}), which supports Boyer's energy argument that there is a delicate balance between the vacuum field driving and the radiation damping. Overall, the RED harmonic oscillator is distinct from the classical harmonic oscillator in that its oscillation amplitude modulates at the time scale of coherence time $\tau_{coh}$ (\Fig{fig:CMREDfinal}), and the resulting RED probability distribution can be thought of as a sum of classical double-peak distributions with a specific amplitude distribution (\Fig{fig:amp}). 

\section{Discussion---Application of RED Simulation to Other Physical Systems}\label{sec:Discussion}

	In quantum mechanics, the harmonic oscillator has excited, coherent, and squeezed states. A natural extension of our current work is to search for the RED correspondence of such states. Currently, we are investigating how a Gaussian pulse with different harmonics of $\wo$ will affect the RED harmonic oscillator. Can the RED harmonic oscillator support a discrete absorption spectrum, and if so, how does it compare with the prediction from quantum mechanics? Such a study is interesting in the broader view of Milonni's comment that RED is unable to account for the discrete energy levels of interacting atoms \cite{Milonni}, and Boyer's comment that at present the line spectra of atoms is still unexplained in RED \cite{Boyer:review}. 
	
	The current method of RED simulation may also be applicable to study other quantum systems that are related to the harmonic oscillator, such as a charged particle in a uniform magnetic field and the anharmonic oscillator \cite{Boyer:diamagnetism,Boyer:anharmonic}. For the first example, classically, a particle in a uniform magnetic field performs cyclotron motion. Such a system can be viewed as a two-dimensional oscillator, having the natural frequency set by the Larmor frequency. On the other hand, a quantum mechanical calculation for the same system reveals Landau quantization. The quantum orbitals of cyclotron motion are discrete and degenerate. Such a system presents a challenge to RED. For the second example, a harmonic potential can be modified to include anharmonic terms of various strength. Heisenberg considered such a system a critical test in the early development of quantum mechanics \cite{Gottfried,Aitchison}. We think that a study of the anharmonic oscillator is thus a natural extension of our current study and may serve as a test for RED. 
	
	Lastly, over the last decades there has been a sustained interest to explain the origin of electron spin and the mechanism behind the electron doubleslit diffraction with RED \cite{de la Pena, Sachidanandam, Kracklauer, Cavalleri}. Several attempts were made to construct a dynamical model that accounts for electron spin. In 1982, de la Pe\~{n}a calculated the phase averaged mechanical angular momentum of a three-dimensional harmonic oscillator. The result deviates from the electron spin magnitude by a factor of 2 \cite{de la Pena}. One year later, Sachidanandam derived the intrinsic spin one-half of a free electron in a uniform magnetic field \cite{Sachidanandam}. Whereas Sachidanandam's calculation is based on the phase averaged canonical angular momentum, his result is consistent with Boyer's earlier work where Landau diamagnetism is derived via the phase averaged mechanical angular momentum of a free electron in a uniform magnetic field\cite{Boyer:diamagnetism}. Although these results are intriguing, the most important aspect of spin, the spin quantization, has not been shown. If passed through a Stern-Gerlach magnet, will the electrons in the RED description split into two groups of trajectories\footnote{
Electron Stern-Gerlach effect is an interesting but controversial topic in its own right. Whereas Bohr and Pauli asserted that an electron beam can not be separated by spin based on the concept of classical trajectories \cite{Pauli}, Batelaan and Dehmelt argue that one can do so with certain Stern-Gerlach-like devices \cite{Batelaan: stern-gerlach1, Batelaan: stern-gerlach2,Dehmelt1,Dehmelt2}.}? At this point, the dynamics becomes delicate and rather complex. To further investigate such a model of spin, an RED simulation may be helpful.

\begin{figure}
\centering
\scalebox{0.55}{\includegraphics{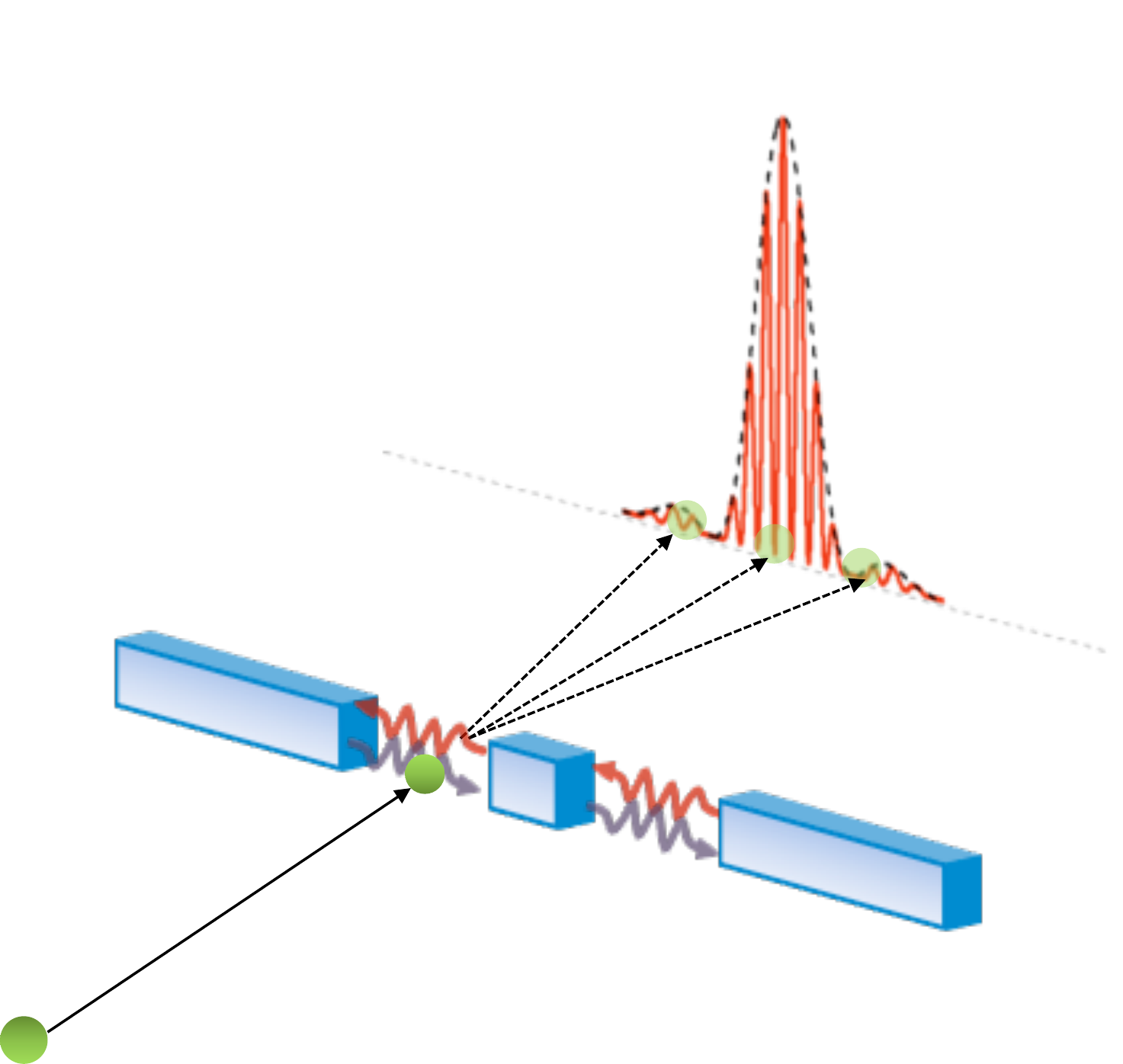}}
\caption{Schematic illustration of RED doubleslit diffraction. Several authors have proposed a RED (or SED) mechanism that explains the electron doubleslit diffraction \cite{Boyer:RED, Kracklauer,Cavalleri}. A central idea provided by RED is that the modes of a vacuum field in one slit are determined by the presence of both slits. This idea allows the superposition principle to be merged with the concept of particle trajectories.}
\label{fig:d-slit} 
\end{figure}

	On the other hand, over the years claims have been made that RED may predict doubleslit electron diffraction \cite{Boyer:RED, Kracklauer,Cavalleri}. In order to explain the experimentally observed electron doubleslit diffraction\footnote{
See \cite{Bach} for a movie of the single electron buildup of a doubleslit diffraction pattern.} \cite{Jonsson1,Jonsson2,Pozzi}, different mechanisms motivated by RED were proposed \cite{Kracklauer,Cavalleri}, while a detailed calculation of the diffracting vacuum field in the vicinity of a slit structure has been given \cite{Avendano}. In 1999, Kracklauer suggested that particles steered by the modulating waves of the RED vacuum field should display a diffraction pattern when passing through a slit, since the RED vacuum field itself is diffracted \cite{Kracklauer}. In recent years, another diffraction mechanism is proposed by Cavalleri \textit{et al.}\ in relation to a postulated electron spin motion \cite{Cavalleri}. Despite of these efforts, Boyer points out in a recent review article that at present there is still not a concrete RED calculation on the doubleslit diffraction \cite{Boyer:review}. Boyer suggests that as the correlation function of the RED vacuum field near the slits is modified by the slit boundary, the motion of the electron near the slits should be influenced as well. Can the scattering of the RED vacuum field be the central mechanism behind the electron doubleslit diffraction (\Fig{fig:d-slit})? As Heisenberg's uncertainty relation is a central feature in all matter diffraction phenomena, any proposed mechanism for electron doubleslit diffraction must be able to account for Heisenberg's uncertainty relation. This is exactly what RED does for the harmonic oscillator. Thus, we hope that the current simulation method may help providing a detailed investigation for any proposed RED diffraction mechanism.

\begin{acknowledgements}
This work was completed utilizing the Holland Computing Center of the University of Nebraska. Especially, the authors thank Dr.~Adam~Caprez, Charles~Nugent, Praga~Angu, and Stephen~Mkandawire for their work on parallelizing the RED simulation code. Huang thanks Dr.~Bradley~A.~Shadwick, Roger~Bach and Dr.~Steven~R.~Dunbar for stimulating discussions. This work is supported by NSF Grant No.~0969506.
\end{acknowledgements}

\appendix

\section{RED Vacuum Field in Unbounded and Bounded Space}
	In ``unbounded'' space, the modes are continuous and the field is expressed in terms of an integral. In ``bounded'' space, the modes are discrete and the field is expressed in terms of a summation. In both cases, the expression for the field amplitude needs to be obtained (see \ref{app:FreeSpaceField} and \ref{app:BoundedSpaceField}). The integral expression helps comparison with analytical calculations in previous papers \cite{Boyer:RED, Boyer:diamagnetism, Boyer:unrah, Boyer:casimir, Boyer:RED&QM}, while the summation expression is what we use in our numerical work.

\subsection{Unbounded Space}\label{app:FreeSpaceField}
	The homogeneous solution of Maxwell's equations in unbounded space is equivalent to the solution for a wave equation,
\begin{equation}
\begin{split}
\vect{E}(\vect{r},t) =& \frac{1}{2}  \sum_{\lambda=1}^2 \int \! d^3k \; \pol(\mode) \left(\tilde{A}(\mode)e^{i(\kdr-\wt)} \right. \\
			      & \left. + \tilde{A}^{\ast}(\mode)e^{-i(\kdr-\wt)} \right) \\
\vect{B}(\vect{r},t) =& \frac{1}{2c}  \sum_{\lambda=1}^2 \int \! d^3k \; ( {\hat{\veck} \times \pol(\mode)} ) \left(\tilde{A}(\mode)e^{i(\kdr-\wt)} \right.\\
			      & \left. + \tilde{A}^{\ast}(\mode)e^{-i(\kdr-\wt)}\right),
\end{split}
\end{equation}
where $\tilde{A}(\mode)$ is the undetermined field amplitude for the mode $(\mode)$ and has the unit of electric field (V/m), $\hat{\veck}$ is defined as the unit vector of $\veck$, and the two vectors, $\pol(\veck,1)$ and $\pol(\veck,2)$, describe an orthonormal polarization basis in a plane that is perpendicular to the wave vector $\veck$.

	Without loss of generality, a random phase \disp{e^{i\rphase}} can be factored out from the field amplitude \disp{\tilde{A}(\mode) = A(\mode)e^{i\rphase}},
\begin{equation}\label{freefield1}
\begin{split}
\vect{E}(\vect{r},t) =& \frac{1}{2}  \sum_{\lambda=1}^2 \int \! d^3k \; \pol(\mode) \left(A(\mode)e^{i(\kdr-\wt)}e^{i\rphase} \right. \\
			      & \left. + A^{\ast}(\mode)e^{-i(\kdr-\wt)}e^{-i\rphase} \right) \\
\vect{B}(\vect{r},t) =& \frac{1}{2c}  \sum_{\lambda=1}^2 \int \! d^3k \; \Bpol(\mode) \left(A(\mode)e^{i(\kdr-\wt)}e^{i\rphase} \right. \\
			      & \left. + A^{\ast}(\mode)e^{-i(\kdr-\wt)}e^{-i\rphase} \right).
\end{split}
\end{equation}
where $\Bpol(\mode) \equiv {\hat{\veck} \times \pol(\mode)}$. The field amplitude $A(\mode)$ can be determined through the phase averaged energy density,
\begin{equation}
\ravg{u} = \frac{\eps}{2}\ravg{\left| \vect{E} \right|^2} + \frac{1}{2\mu_0}\ravg{\left| \vect{B} \right|^2},
\end{equation}
where $\eps$ is the vacuum permittivity, $\mu_0$ is the vacuum permeability, and $\tilde{\theta}$ is the random phase in \Eq{freefield1}. To evaluate the phase averaged energy density $\ravg{u}$, we first calculate $\left| \vect{E} \right|^2$ and $\left| \vect{B} \right|^2$ using \Eq{freefield1},
\begin{equation}\label{freefield2}
\begin{split}
\left| \vect{E} \right|^2(\vect{r},t)
&= \vect{E}(\vect{r},t) \vect{E}^{\ast}(\vect{r},t) \\
&= \frac{1}{4} \sum_{\lambda,\lambda'} \int \!\! d^3k \!\! \int \!\! d^3k' \; (\pol(\mode)\cdot\pol(\modep))f, \\			  
\left| \vect{B} \right|^2(\vect{r},t)
&= \vect{B}(\vect{r},t) \vect{B}^{\ast}(\vect{r},t) \\
&= \frac{1}{4c^2} \sum_{\lambda,\lambda'} \int \!\! d^3k \!\! \int \!\! d^3k' \; (\Bpol(\mode)\cdot\Bpol(\modep))f,
\end{split}
\end{equation}
where $f \equiv f(\mode;\modep;\vect{r},t)$,
\begin{equation}\label{freefield3}
\begin{split}
&f(\mode;\modep;\vect{r},t) \\
&= A(\mode)A^{\ast}(\modep)e^{i(\kdr-\wt+\rphase)}e^{-i(\kdrp-\wtp+\rphasep)} \\ 
&\quad +A(\mode)A(\modep)e^{i(\kdr-\wt+\rphase)}e^{i(\kdrp-\wtp+\rphasep)} \\
&\quad +A^{\ast}(\mode)A^{\ast}(\modep)e^{-i(\kdr-\wt+\rphase)}e^{-i(\kdrp-\wtp+\rphasep)} \\ 				 
&\quad +A^{\ast}(\mode)A(\modep)e^{-i(\kdr-\wt+\rphase)}e^{i(\kdrp-\wtp+\rphasep)}.		 
\end{split}
\end{equation}
The random phase average can be calculated with the following relation \cite{Boyer:RED&QM},
\begin{equation}\label{freefield4}
\left\{
\begin{array}{l}
\ravg{e^{\pm i(\rphase+\rphasep)}} = 0, \\
\\
\ravg{e^{\pm i(\rphase-\rphasep)}} = \delta_{\lambda',\lambda}\delta^3(\veck'-\veck). \\
\end{array}
\right.
\end{equation}
Applying \Eq{freefield4} to \Eq{freefield3}, we obtain
\begin{equation}\label{freefield4-2}
\begin{split}
&\ravg{f(\mode;\modep;\vect{r},t)} \\
&= A(\mode)A^{\ast}(\modep)e^{i(\kdr-\wt)}e^{-i(\kdrp-\wtp)}\delta_{\lambda',\lambda}\delta^3(\veck'-\veck) \\ 
&\quad + A^{\ast}(\mode)A(\modep)e^{-i(\kdr-\wt)}e^{i(\kdrp-\wtp)}\delta_{\lambda',\lambda}\delta^3(\veck'-\veck).		 
\end{split}
\end{equation}
Consequently, $\ravg{\left| \vect{E} \right|^2}$ and $\ravg{\left| \vect{B} \right|^2}$ can be evaluated using Eqs.~(\ref{freefield2}) and (\ref{freefield4-2}),
\begin{equation}
\begin{split}
\ravg{\left| \vect{E} \right|^2} 
&= \frac{1}{4} \sum_{\lambda,\lambda'} \int \!\! d^3k \!\! \int \!\! d^3k' \; (\pol(\mode)\cdot\pol(\modep)) \ravg{f} \\	
&= \frac{1}{2}  \sum_{\lambda=1}^2 \int \! d^3k \; \left| A(\mode) \right|^2, \\
\ravg{\left| \vect{B} \right|^2} 
&= \frac{1}{4c^2} \sum_{\lambda,\lambda'} \int \!\! d^3k \!\! \int \!\! d^3k' \; (\Bpol(\mode)\cdot\Bpol(\modep)) \ravg{f} \\
&= \frac{1}{2c^2}  \sum_{\lambda=1}^2 \int \! d^3k \; \left| A(\mode) \right|^2.
\end{split}
\end{equation}
The above calculation leads to a relation between the field amplitude $A(\mode)$ and the phase averaged energy density $\ravg{u}$ in unbounded space,
\begin{equation}\label{freefield5}
\ravg{u} = \frac{\eps}{2} \sum_{\lambda=1}^2 \int \! d^3k \; \left| A(\mode) \right|^2.
\end{equation}
Now, if we postulate that the vacuum energy is $\hbar\omega/2$ for each mode $(\mode)$, then in a bounded cubic space of volume $V$ the vacuum energy density is
\begin{equation}
\begin{split}
u_{vac} &= \frac{1}{V}\sum_{\mode}\frac{\hbar\omega}{2} \\
	     &= \frac{1}{V}\sum_{\mode} \frac{\hbar\omega}{2} \left( \frac{L_x}{2\pi}\Delta k_x \right) \left( \frac{L_y}{2\pi}\Delta k_y \right)\left( \frac{L_z}{2\pi}\Delta k_z \right) \\
	     &= \sum_{\lambda=1}^2 \frac{1}{(2\pi)^3} \sum_{\veck} \Delta^3k \frac{\hbar\omega}{2}.
\end{split}
\end{equation}
In the limit of unbounded space (i.e.\ $V \rightarrow \infty$), the volume element  $\Delta^3k$ becomes differential (i.e.\ $\Delta^3k \rightarrow d^3k$) and the vacuum energy density becomes
\begin{equation}
u_{vac} = \sum_{\lambda=1}^2 \frac{1}{(2\pi)^3} \int \! d^3k \frac{\hbar\omega}{2}. 
\end{equation}
Comparing this result with \Eq{freefield5}, we find
\begin{equation}
 \left| A_{vac}(\mode) \right|^2 = \frac{\hbar\omega}{(2\pi)^3\eps}.
\end{equation}
Assuming $A_{vac}(\mode)$ is a positive real number, the vacuum field amplitude in unbounded space is determined,
\begin{equation}
A_{vac}(\mode) = \sqrt{\frac{\hbar\omega}{8\pi^3\eps}}.
\end{equation}
Therefore, the RED vacuum field in unbounded space is found to be
\begin{equation}
\begin{split}
\vect{E}_{vac}(\vect{r},t) =& \frac{1}{2}  \sum_{\lambda=1}^2 \int \! d^3k \; \pol(\mode) \sqrt{\frac{\hbar\omega}{8\pi^3\eps}} \left(e^{i(\kdr-\wt)}e^{i\rphase} \right. \\
				       & \left. + e^{-i(\kdr-\wt)}e^{-i\rphase} \right) \\
\vect{B}_{vac}(\vect{r},t) =& \frac{1}{2c}  \sum_{\lambda=1}^2 \int \! d^3k \; \Bpol(\mode) \left(\sqrt{\frac{\hbar\omega}{8\pi^3\eps}} e^{i(\kdr-\wt)}e^{i\rphase} \right. \\
				       & \left. + e^{-i(\kdr-\wt)}e^{-i\rphase} \right),
\end{split}
\end{equation}
which is \Eq{Boyer:vacE}.

\subsection{Bounded Space}\label{app:BoundedSpaceField}
	The solution of homogeneous Maxwell's equations in bounded space has the summation form,
\begin{equation}
\begin{split}
&\vect{E}(\vect{r},t) = \frac{1}{2} \sum_{\mode} \left(\tilde{A}_{\submode}e^{i(\kdr-\wt)} + \tilde{A}^{\ast}_{\submode}e^{-i(\kdr-\wt)}\right) \pol_{\submode}, \\
&\vect{B}(\vect{r},t) = \frac{1}{2c} \sum_{\mode} \left(\tilde{A}_{\submode}e^{i(\kdr-\wt)} + \tilde{A}^{\ast}_{\submode}e^{-i(\kdr-\wt)}\right) \Bpol_{\submode},
\end{split}
\end{equation}
where $\Bpol_{\submode} = \hat{\veck} \times \pol_{\submode}$, $\tilde{A}_{\submode}$ is the undetermined field amplitude for the mode $(\mode)$ and has the unit of electric field (V/m), $\hat{\veck}$ is defined as the unit vector of $\veck$, and the two vectors, $\pol_{_{\veck,1}}$ and $\pol_{_{\veck,2}}$, describe an orthonormal polarization basis in a plane that is perpendicular to the wave vector $\veck$.

	Using the relation\footnote{
If the two modes are not identical (i.e.\ $\veck' \ne \veck$ or $\lambda' \ne \lambda$), then $e^{i\theta_{\mode}}$ and $e^{i\theta_{\modep}}$ are independent, which leads to the factorization ${\langle{e^{i(\theta_{\mode}\pm \theta_{\modep})}}\rangle}_{\tilde{\theta}} = {\langle{e^{i\theta_{\mode}}}\rangle}_{\tilde{\theta}} {\langle{e^{\pm i\theta_{\modep}}}\rangle}_{\tilde{\theta}} = 0$.}
\begin{equation}
\left\{
\begin{array}{l}
\ravg{e^{\pm i(\subrphase+\subrphasep)}} = 0\\
\\
\ravg{e^{\pm i(\subrphase-\subrphasep)}} = \delta_{\lambda',\lambda}\delta_{\veck',\veck}, \\
\end{array}
\right.
\end{equation}
we can follow the same argument in \ref{app:FreeSpaceField} and obtain the phase averaged energy density in bounded space
\begin{equation}\label{boundedfield1}
\ravg{u} = \frac{\eps}{2} \sum_{\mode} \left| A_{\submode} \right|^2,
\end{equation}
where $\tilde{A}_{\submode}=A_{\submode}e^{i\subrphase}$. Again, if we postulate that the vacuum energy is $\hbar\omega/2$ for each mode $(\mode)$, then in a bounded space of volume $V$ the vacuum energy density is
\begin{equation}\label{boundedfield2}
u_{vac} = \frac{1}{V}\sum_{\mode}\frac{\hbar\omega}{2}.
\end{equation}
Comparing \Eq{boundedfield1} and \Eq{boundedfield2}, the vacuum field amplitude in bounded space is determined,
\begin{equation}
{A_{vac}}_{\submode} = \sqrt{\frac{\hbar\omega}{\eps V}}.
\end{equation}
Therefore, the RED vacuum field in bounded space is
\begin{equation}
\begin{split}
&\vect{E}_{vac}(\vect{r},t) = \frac{1}{2} \sum_{\mode} \left(\sqrt{\frac{\hbar\omega}{\eps V}}e^{i(\kdr-\wt)}e^{i\subrphase} + c.c. \right) \pol_{\submode}, \\
&\vect{B}_{vac}(\vect{r},t) = \frac{1}{2c} \sum_{\mode} \left(\sqrt{\frac{\hbar\omega}{\eps V}}e^{i(\kdr-\wt)}e^{i\subrphase} + c.c \right) \Bpol_{\submode}.
\end{split}
\end{equation}

\section{Isotropic Polarization Vectors}\label{app:isotropy}
A wave vector chosen along the $z$-axis,
\begin{equation}
\tilde{\veck} = 
\begin{pmatrix}
\tilde{k}_x \\ \tilde{k}_y \\ \tilde{k}_z
\end{pmatrix}
=
\begin{pmatrix}
0 \\ 0 \\ k
\end{pmatrix}
,
\end{equation}
has an orthonormal polarization basis in the $xy$-plane,
\begin{equation}
\tilde{\pol}_{_{\veck,1}} = 
\begin{pmatrix}
\cos{\chi} \\ \sin{\chi} \\ 0
\end{pmatrix}
,
\tilde{\pol}_{_{\veck,2}} = 
\begin{pmatrix}
-\sin{\chi} \\ \cos{\chi} \\ 0
\end{pmatrix}
,
\end{equation}
where the random angle $\chi$ is uniformly distributed in $[0,2\pi]$. To obtain the wave vector $\veck$ in \Eq{mslct:layer1}, $\tilde{\veck}$ can be first rotated counterclockwise about the $y$-axis by an angle $\theta$, then counterclockwise about the $z$-axis by an angle $\phi$. The corresponding rotation matrix is described by
\begin{equation}
\begin{split}
\hat{R} &= \hat{R}_{\phi}^{(z)}\hat{R}_{\theta}^{(y)} \\
&=
\begin{pmatrix}
\cos{\phi}  & -\sin{\phi}  & 0 \\
\sin{\phi}   & \cos{\phi} & 0 \\
0		& 0		   & 1 
\end{pmatrix}
\begin{pmatrix}
\cos{\theta}   & 0 & \sin{\theta} \\
0		    & 1 & 0  \\
-\sin{\theta}  & 0 & \cos{\theta} 
\end{pmatrix}
\\
&=
\begin{pmatrix}
\cos{\theta}\cos{\phi} &-\sin{\phi} &\cos{\phi}\sin{\theta} \\
\cos{\theta}\sin{\phi}  &\cos{\phi} &\sin{\phi}\sin{\theta}  \\
-\sin{\theta}		& 0		  & \cos{\theta} 		     
\end{pmatrix}
,
\end{split}
\end{equation}
and $\veck$ is obtained accordingly,
\begin{equation}
\veck = \hat{R}\tilde{\veck} = 
\begin{pmatrix}
k\sin{\theta}\cos{\phi} \\ k\sin{\theta}\sin{\phi} \\ k\cos{\theta}
\end{pmatrix}. 
\end{equation}
In the same manner, we can rotate $\tilde{\pol}_{_{\veck,1}}$ and $\tilde{\pol}_{_{\veck,2}}$ with the rotation matrix $\hat{R}$, and obtain an isotropically distributed\footnote{
After the rotation, the uniformly distributed circle will span into a uniformly distributed spherical surface.} polarization basis as described in \Eq{pol}, 
\begin{equation}
\begin{split}
&\pol_{_{\veck,1}} = \hat{R}\tilde{\pol}_{_{\veck,1}} =
\begin{pmatrix}
\cos{\theta}\cos{\phi}\cos{\chi}-\sin{\phi}\sin{\chi} \\
\cos{\theta}\sin{\phi}\cos{\chi}+\cos{\phi}\sin{\chi} \\
-\sin{\theta}\cos{\chi}
\end{pmatrix}
\\
&\pol_{_{\veck,2}} = \hat{R}\tilde{\pol}_{_{\veck,2}} = 
\begin{pmatrix}
-\cos{\theta}\cos{\phi}\sin{\chi}-\sin{\phi}\cos{\chi} \\
-\cos{\theta}\sin{\phi}\sin{\chi}+\cos{\phi}\cos{\chi} \\
\sin{\theta}\sin{\chi}
\end{pmatrix}
.
\end{split}
\end{equation}

\section{Repetitive Time}\label{app:repetition}

\begin{figure}[t]
\centering
\scalebox{0.45}{\includegraphics{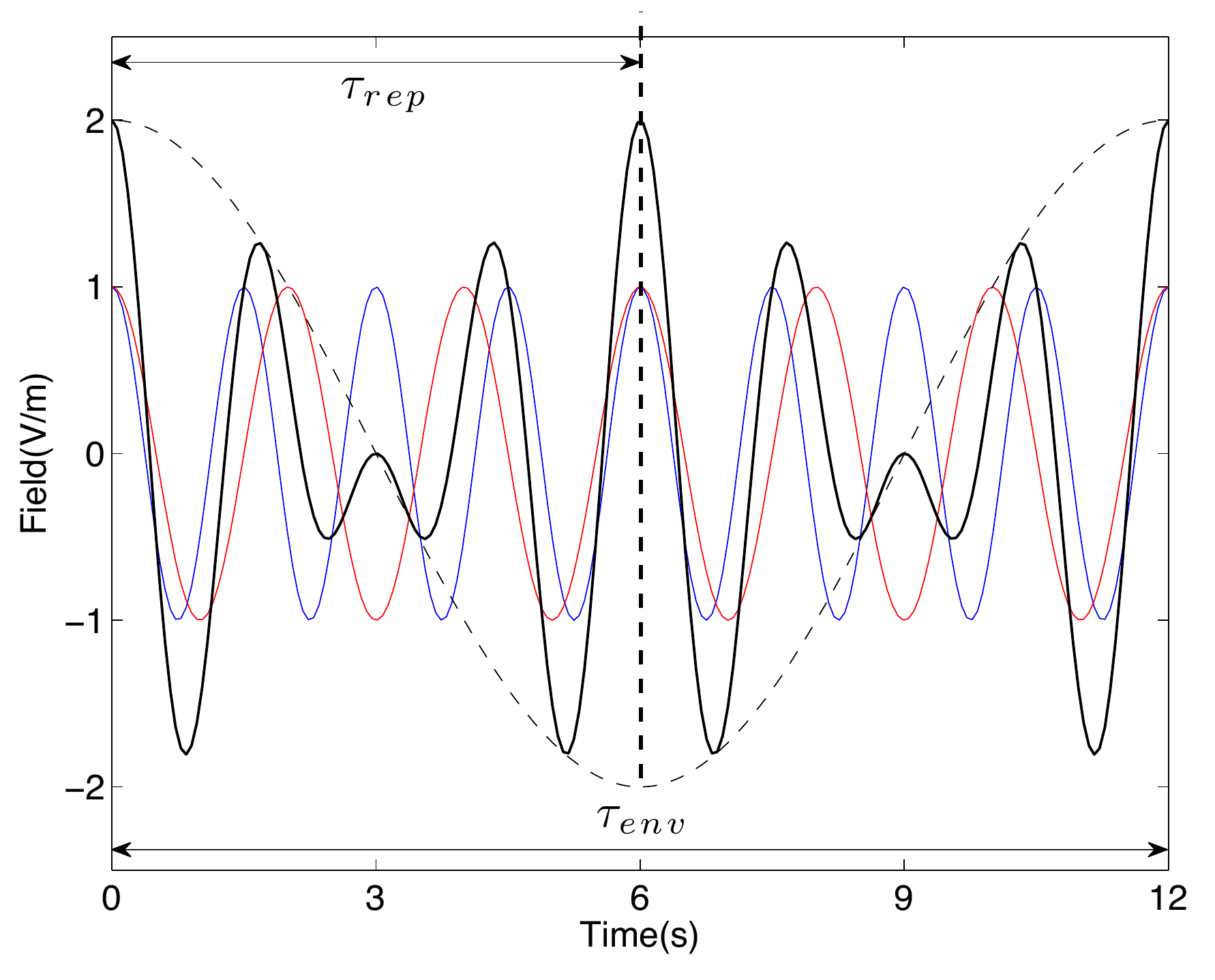}}
\caption{Repetition time of a beat wave. A beat wave (black solid line) is made of two frequency components (red \& blue solid line). The oscillation periods of the two frequency components are  $T_1 = 1.5$ and $T_2 = 2$, thus the repetition time is $\tau_{rep} = [T_1,T_2]_{LCM} = 6$. Notice that the periodicity of the envelope (black dash line) is $\tau_{env} = 12$, which is different from the repetition time $\tau_{rep} = 6$.}
\label{fig:rep} 
\end{figure}

A field composed of finite discrete frequencies, 
\begin{equation}
E(t) = \sum^{N}_{k=1} E_k \cos{(\omega_k t)},
\end{equation}
repeats itself at the least common multiple (LCM) of all the periods of its frequency components, 
\begin{gather}
E(t+\tau_{rep}) = E(t), \\
\tau_{rep} = [T_1,T_2,\ldots,T_N]_{LCM}, \label{rep_1_1}
\end{gather}
where \disp{T_k = \frac{2\pi}{\omega_k}}. An example of a two-frequency beat wave is given in \Fig{fig:rep}.

Given the frequency spectrum of $E(t)$, one can draw a relation between the repetition time $\tau_{rep}$ and the greatest common divider (GCD) of the frequencies, 
\begin{equation}\label{rep1}
\tau_{rep} = \frac{2\pi}{(\omega_1,\omega_2,\ldots,\omega_N)_{GCD}}.
\end{equation}
The derivation of the relation in \Eq{rep1} is the following. First, we factorize the sum of all the frequencies into two terms, 
\begin{equation}
\begin{split}
&\omega_1 + \omega_2 + \ldots + \omega_N \\
&= \frac{2\pi}{T_1} + \frac{2\pi}{T_2} + \ldots + \frac{2\pi}{T_N} \\
& =  \frac{2\pi}{[T_1,T_2,\ldots,T_N]_{LCM}} (n_1 + n_2 + \ldots + n_N),
\end{split}
\end{equation}
where $n_k$ are positive integers, 
\begin{equation}\label{rep2}
n_k = \frac{[T_1,T_2,\ldots,T_N]_{LCM}}{T_k}.
\end{equation}
Now, it is true that
\begin{equation}
(n_1, n_2, \ldots, n_N)_{GCD} = 1, 
\end{equation}
because the reverse, $(n_1, n_2, \ldots, n_N)_{GCD} > 1$, would lead to a contradiction to \Eq{rep2}. Therefore, one can conclude that
\begin{equation}\label{rep_1_2}
\begin{split}
\dw_{gcd} &\equiv (\omega_1,\omega_2,\ldots,\omega_N)_{GCD} \\
		& = \frac{2\pi}{[T_1,T_2,\ldots,T_N]_{LCM}}.
\end{split}
\end{equation}
From \Eq{rep_1_1} and \Eq{rep_1_2}, the relation in \Eq{rep1}
\begin{equation}
\tau_{rep} = \frac{2\pi}{\dw_{gcd}}
\end{equation}
is drawn. 

Since the simulation should only be carried through an integration time $\tau_{int} \le \tau_{rep}$ to avoid repetitive solutions, in our case the choice of the integration time (\Eq{rep})
\begin{equation}\label{rep3}
\tau_{int} = \frac{2\pi}{\dw}, 
\end{equation}
where $\dw$ is the smallest frequency gap ($\dw \le \dw_{gcd}$), suffices our purpose. The frequency gap as a function of $\kappa$ can be estimated using \Eq{mslct:layer2},
\begin{equation}
\begin{split}
\dw(\kappa) &= \omega(\kappa) - \omega(\kappa-\dkapa) \\
       &= c(3\kappa)^{1/3} - c(3\kappa)^{1/3} \left( 1-\frac{\dkapa}{\kappa}\right)^{1/3}.
\end{split}
\end{equation}
Applying the sharp resonance condition (\Eq{sharpreson}) to Eqs.~(\ref{dka}) and (\ref{mslct:layer3}), it can be further shown that $\dkapa$ is much smaller than $\kappa_0$ and $\kappa \simeq \kappa_0$,
\begin{gather}
\frac{\dkapa}{\kappa_0} = \left(\frac{3}{N_{\veck}-1}\right) \frac{\Delta}{\wo} \ll 1, \\
\kappa = \kappa_0 + O \left( \frac{\Delta}{\wo} \right),
\end{gather}
where \disp{\kappa_0 = \frac{1}{3} \left( \frac{\wo}{c} \right)^3}. Therefore, the size of $\dw(\kappa)$ is approximately fixed within the sampled frequency range $\Delta$, and the smallest frequency gap $\dw$ can be approximated as
\begin{equation}
\dw \simeq \frac{c(3\kappa_0)^{1/3}}{3}\frac{\dkapa}{\kappa_0}.
\end{equation}

\end{document}